\documentclass[aps,prc,preprint,showpacs,showkeys,superscriptaddress,amsmath,amssymb,floatfix,nofootinbib]{revtex4}

\usepackage[dvips]{graphicx,color}
\usepackage{color}
\usepackage{dcolumn}
\usepackage{bm}

\def\pmb#1{\setbox0=\hbox{#1}%
  \kern-.025em\copy0\kern-\wd0 
  \kern.05em\copy0\kern-\wd0
  \kern-.025em\raise.0433em\box0 }

\def\lambdabar{\protect\@lambdabar}
\def\@lambdabar{%
\relax
\bgroup
\def\@tempa{\hbox{\raise.73\ht0
\hbox to0pt{\kern.25\wd0\vrule width.5\wd0
height.1pt depth.1pt\hss}\box0}}%
\mathchoice{\setbox0\hbox{$\displaystyle\lambda$}\@tempa}%
{\setbox0\hbox{$\textstyle\lambda$}\@tempa}%
{\setbox0\hbox{$\scriptstyle\lambda$}\@tempa}%
{\setbox0\hbox{$\scriptscriptstyle\lambda$}\@tempa}%
\egroup
}

\begin{document}

\preprint{J-PARC-TH-0167}

\title{\boldmath
Production of a $^{4}_\Lambda$He hypernucleus in the $^4$He($\pi$,~$K$) reactions reexamined
}

\author{Toru~Harada}%
\email{harada@osakac.ac.jp}
\affiliation{%
Center for Physics and Mathematics,
Osaka Electro-Communication University, Neyagawa, Osaka, 572-8530, Japan
}
\affiliation{%
J-PARC Branch, KEK Theory Center, Institute of Particle and Nuclear Studies,
High Energy Accelerator Research Organization (KEK),
203-1, Shirakata, Tokai, Ibaraki, 319-1106, Japan
}

\author{Yoshiharu~Hirabayashi}%
\affiliation{%
Information Initiative Center, 
Hokkaido University, Sapporo, 060-0811, Japan
}

\date{\today}

\begin{abstract}
We investigate theoretically production cross sections of the $J^\pi=$ 0$^+$ ground state 
of a $^4_\Lambda$He hypernucleus in the $^4$He($\pi$,~$K$) reaction
with a distorted-wave impulse approximation using the optimal Fermi-averaged 
$\pi N\to K\Lambda$ $t$ matrix.  
We demonstrate the sensitivity of the production cross sections to the $\Lambda$ wavefunctions 
obtained from $3N$-$\Lambda$ potentials and to meson distorted waves in eikonal distortions.
It is shown that the calculated laboratory cross sections of the 0$^+$ ground state in $^4_\Lambda$He
amount to $d\sigma/d\Omega_{\rm lab} \simeq$ 11 $\mu$b/sr at $p_{\pi}=$ 1.05 GeV/$c$ 
in the $K$ forward direction because of an advantage of the use of the s-shell target nucleus such as $^4$He.
The importance of the recoil effects and the energy dependence of the $\pi N\to K\Lambda$ 
cross sections is also discussed. 
\end{abstract}
\maketitle


\section{Introduction}

Recently, unexpected short lifetimes of a ${^{3}_\Lambda{\rm H}}$ hypernucleus were measured 
in hypernuclear production of high-energy heavy-ion collisions \cite{HypHI13,ALICE16,STAR18}; 
the world average lifetime of $\tau^{\rm (av)}({^3_\Lambda{\rm H}})=$ $185^{+23}_{-28}$ ps 
is shorter than the free lifetime 
$\tau_\Lambda=$ $263.2 \pm 2.0$ ps by about 30\%.
However, ALICE Collaboration \cite{ALICE18} newly reported the result of 
$\tau({^3_\Lambda{\rm H}})=$ $237^{+33}_{-36}$ ps which is moderately closer to $\tau_\Lambda$. 
This is one of the most topical issues to study hypernuclear physics \cite{Gal19}, 
and is essential because $^3_\Lambda$H is the lightest hypernucleus which 
provides valuable information on interacting a $\Lambda$ hyperon with nucleons \cite{Kamada98}. 
In addition, HypHI Collaboration \cite{HypHI13} suggested that 
the lifetime of a $^4_\Lambda$H hypernucleus $\tau(^4_\Lambda{\rm H})=$ $140^{+48}_{-33}$ ps 
is shorter than $\tau(^4_\Lambda{\rm H})=$ $194^{+28}_{-26}$ ps measured in 
stopped $K^-$ experiments at KEK \cite{Outa98}, 
whereas theoretical calculations \cite{Motoba92,Fuse95} predicted 
$\tau(^4_\Lambda{\rm H})=$ 196--264 ps which depend on the $\Lambda$ wavefunctions. 
To solve the lifetime puzzle, experimental measurements of the 
${^{3,4}_\Lambda{\rm H}}$ lifetime have been planned in ($K^-$,~$\pi^0$) and ($\pi^-$,~$K^0$) reactions 
on $^{3,4}$He targets at J-PARC \cite{J-PARC-P73,J-PARC-P74}.
Therefore, it is important to investigate theoretically production of $A=$ 3, 4 hypernuclei 
via the ($\bar{K}$,~$\pi$) and ($\pi$,~$K$) reactions on $^{3,4}$He targets \cite{Harada19}. 
Especially, we believe that $\Lambda$ production studies on the $^{4}$He target 
are useful to settle mysterious problems related to the $A=4$ hypernuclei \cite{Gal16},
e.g., overbinding/underbinding anomaly \cite{Dalitz72,Akaishi00}, 
charge symmetry breaking (CSB) \cite{Dalitz64,Gal15}, and so on.

Many theoretical studies of hypernuclear spectroscopy have been performed
in nuclear ($K^-$,~$\pi^-$), ($\pi^+$,~$K^+$), and ($\gamma$,~$K^+$) reactions \cite{Gal16}. 
Production cross sections of $\Lambda$ hypernuclear states for each reaction 
are usually characterized by a specific momentum transfer to a $\Lambda$ hyperon.
The ($K^-$,~$\pi^-$) reactions on $^4$He \cite{Harada98} can produce  
a $\Lambda$ substitutional state in $^4_\Lambda$He under recoilless conditions. 
On the other hand, the ($\pi^+$,~$K^+$) reactions on $^4$He enable us to give 
large momentum transfers of $q=$ 300--500 MeV/$c$ \cite{Shinmura86}, 
populating a $\Lambda$ stretched state which seems to be disadvantageous to 
the $0s_N \to 0s_\Lambda$ transition in terms of a momentum matching law \cite{Dover80};
however, the production cross sections of $^4_\Lambda$He via the ($\pi^+$,~$K^+$) reaction on $^4$He 
were predicted theoretically \cite{Harada06} for the non-mesonic weak decay of 
$^4_\Lambda$He and $^4_\Lambda$H at J-PARC \cite{Ajimura06}. 
Consequently, it is worth revisiting investigation of $^4_\Lambda$He production 
via the ($\pi$,~$K$) reactions on $^4$He 
because we have now been achieved a good understanding of the nuclear ($\pi$,~$K$) 
reactions \cite{Harada05}. 

In this paper, we investigate theoretically the production cross sections of 
the $J^\pi=$ 0$^+$ ground state ($0^+_{\rm g.s.}$) of a $^4_\Lambda$He hypernucleus 
via the $^4$He($\pi^+$,~$K^+$) reaction 
in the distorted-wave impulse approximation.
We demonstrate the differential cross sections at $p_{\pi}=$ 1.05 GeV/$c$ 
in the $K^+$ forward direction in the laboratory system. 
We discuss medium effects of the $\pi N \to K\Lambda$ amplitude 
in nuclear ($\pi$,~$K$) reactions, and recoil effects in the light hypernucleus 
such as $^4_\Lambda$He.
Our results on $^4_\Lambda$He production via $^4$He($\pi^+$,~$K^+$) reactions  
are expected to be identical to those on $^4_\Lambda$H mirror production via 
$^4$He($\pi^-$,~$K^0$) reactions because charge independence guarantees 
${f}_{\pi^+ n\to K^+\Lambda}=-{f}_{\pi^- p\to K^0\Lambda}$
in the ($\pi$,~$K$) reactions on $T=0$ target nuclei.

\section{Calculations}

\begin{figure}[t]
\begin{center}
   \includegraphics[width=\linewidth]{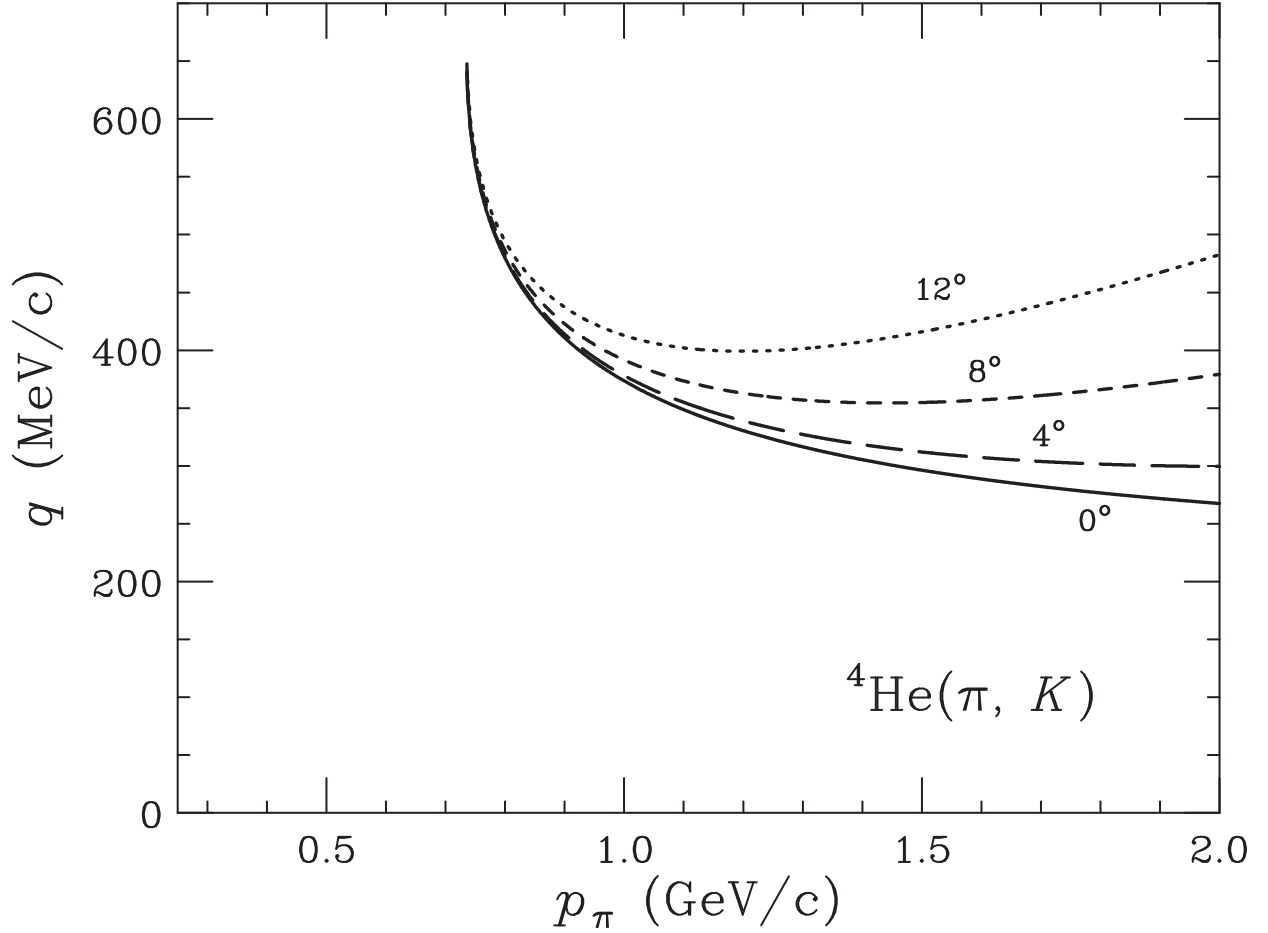}
\end{center}
\caption{\label{fig:1}
Momentum transfer $q$ to the $\Lambda$ final state of $0^+_{\rm g.s.}$ in $^4_\Lambda$He 
for ($\pi$,~$K$) reactions on a $^4$He target at $K$ forward-direction angles 
of $\theta_{\rm lab}=$ $0^\circ$--$12^\circ$ in the lab frame, 
as a function of the incident lab momentum $p_{\pi}$. 
}
\end{figure}

\subsection{Distorted-wave impulse approximation}
\label{sect:2-1}

Let us consider a calculation procedure of hypernuclear production
 for nuclear ($\pi$,~$K$) reactions in the laboratory (lab) frame. 
The double differential cross sections within 
the distorted-wave impulse approximation (DWIA)
\cite{Hufner74,Auerbach83} are given by 
(in units $\hbar=c=1$)
\begin{eqnarray}
   {{d}^2{\sigma} \over {d}E_K {d}\Omega_K}
&=& \beta
    {1 \over {[J_A]}}
    \sum_{m_A}\sum_{B,m_B} \vert\langle {\Psi}_B \vert
    \hat{F} \vert {\Psi}_A \rangle\vert^{2} \nonumber\\
&& \times \delta (E_{K}+E_{B}-E_{\pi}-E_{A}), 
\label{eqn:e1}
\end{eqnarray}
where $[J]=2J+1$, and $E_{K}$, $E_{\pi}$, $E_{B}$ and $E_{A}$ are energies of 
outgoing $K$, incoming $\pi$, 
hypernuclear states and the target nucleus, respectively.
$\Psi_B$ and $\Psi_A$ are wavefunctions of hypernuclear final states 
and the initial state of the target nucleus, respectively. 
The kinematical factor $\beta$ \cite{Tadokoro95,Koike08} 
arising from a translation from a two-body meson-nucleon lab system to a meson-nucleus 
lab system \cite{Dover83} is given by 
\begin{equation}
 \beta=
 \biggl(1+ {E^{(0)}_{K} \over E^{(0)}_{B}}
        {{p^{(0)}_{K} - p^{(0)}_{\pi} \cos\theta_{\rm lab}}
        \over p^{(0)}_{K}} \biggr)
        {p_{K} E_{K} \over p^{(0)}_{K} E^{(0)}_{K}},
\label{eqn:e2}
\end{equation}
where 
$p^{(0)}_{\pi}$ and $p^{(0)}_{K}$ ($E^{(0)}_{K}$ and 
$E^{(0)}_{B}$) are 
lab momenta of $\pi$ and $K$ (lab energies of $K$ and $\Lambda$)
in the two-body $\pi N\to K\Lambda$ reaction, respectively.
Here we considered only the non-spin-flip amplitude because we are 
interested in the $\Delta S=$ 0 cross sections in the $K$ forward direction.
Thus an external operator $\hat{F}$ for the associated production $\pi N \to K \Lambda$ reactions is given by
\begin{eqnarray}
\hat{F}
 &=& \int d{\bm r} \> \chi_{K}^{(-) \ast}({\bm p}_{K},{\bm r})
\chi_{\pi}^{(+)}({\bm p}_{\pi},{\bm r}) \nonumber\\
 &&  \times \sum_{j=1}^A \overline{f}_{\pi N\to K\Lambda}
\delta ({\bm r}-{\bm r}_{j}){\hat O}_j, 
\label{eqn:e3}
\end{eqnarray}
where we assume zero-range interaction for the $\pi N\to K\Lambda$ transitions.
Distorted waves of $\chi_{K}^{(-) \ast}$ and $\chi_{\pi}^{(+)}$ 
are obtained with the help of the eikonal approximation \cite{Hufner74}. 
${\hat O}_j$ is a baryon operator changing 
$j$th nucleon into a $\Lambda$ hyperon in the nucleus, 
and ${\bm r}$ is the relative coordinate between the mesons and 
the center-of-mass (c.m.) of the nucleus; $\overline{f}_{\pi N\to K\Lambda}$
is the Fermi-averaged non-spin-flip amplitude for the $\pi N\to K\Lambda$ 
reactions in nuclei on the lab frame \cite{Harada05}.

\begin{figure}[t]
\begin{center}
  \includegraphics[width=\linewidth]{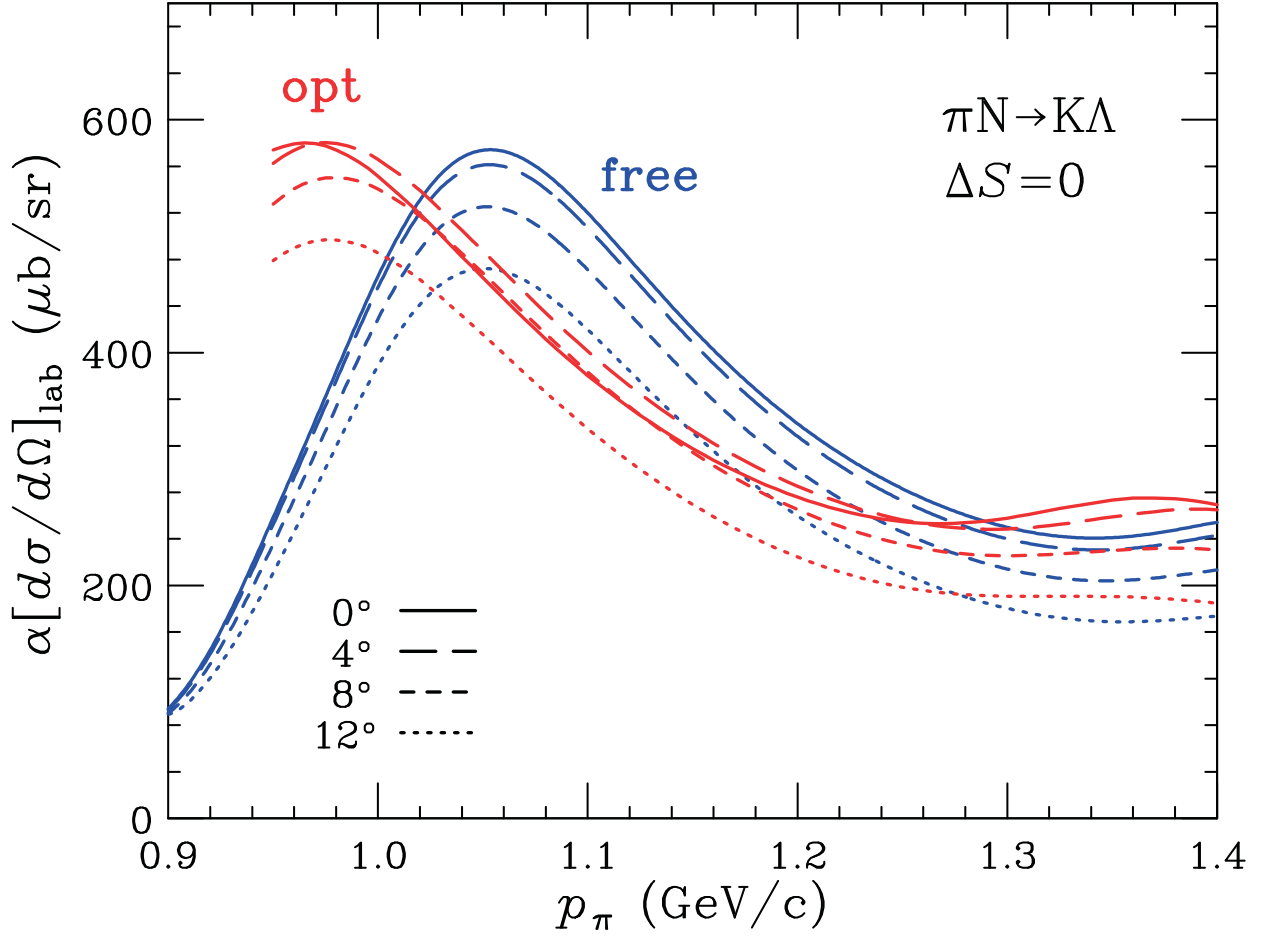}
\end{center}
\caption{\label{fig:2}
Optimal Fermi-averaged $\pi N \to  K\Lambda$ lab cross sections of 
$\alpha \langle d\sigma/d\Omega_K \rangle^{\rm opt}_{\pi N \to K\Lambda}$ 
for the $\Delta S=$ 0 nuclear ($\pi$,~$K$) reactions in nuclei \cite{Harada05}, 
as a function of the incident lab momentum $p_\pi$. 
The kinematics for a $^4$He target and the non-spin-flip amplitude 
at $K$ forward-direction lab angles of $\theta_{\rm lab}$ are used.
The elementary $\pi N \to  K\Lambda$ lab cross sections 
of $\alpha \langle d\sigma/d\Omega_K\rangle^{\rm free}_{\pi N \to K\Lambda}$ 
for $\Delta S=$ 0 in free space \cite{Sotona89} are also drawn.
}
\end{figure}

\begin{table*}[tbh]
\caption{
\label{tab:1}
Kinematical values and the $\pi N\to K \Lambda$ lab cross sections 
in the nuclear ($\pi$,~$K$) reactions on a $^4$He target at $p_{\pi}=$ 1.05 GeV/$c$. 
The non-spin-flip amplitudes in the $\pi N\to K \Lambda$ reaction are used. 
}
\begin{ruledtabular}
\begin{tabular}{cccccccc}
$\theta_{\rm lab}$ 
& $p_K$
& $q$   
& $q_{\rm eff}$\footnotemark[1] 
& $\beta$
& $\alpha$
& \multicolumn{2}{c}
{$\alpha \langle d\sigma/d\Omega \rangle_{\small \pi N\to\Lambda K}$}
\\
\noalign{\smallskip}
 \cline{7-8}  
\noalign{\smallskip}
(degree)
& (MeV/$c$)
& (MeV/$c$)
& (MeV/$c$)
&
&
& free ($\mu$b/sr)
& opt  ($\mu$b/sr) \\
\noalign{\smallskip}\hline\noalign{\smallskip}
  0 &  690.1  &   359.9  &   271.1   &   0.684   &  0.828  &  574  &   464   \\
  4 &  689.5  &   365.4  &   275.3   &   0.686   &  0.829  &  561  &   487   \\ 
  8 &  687.7  &   381.3  &   287.3   &   0.692   &  0.834  &  525  &   468   \\ 
 12 &  684.6  &   406.2  &   306.1   &   0.703   &  0.843  &  472  &   416   \\ 
\end{tabular}
\end{ruledtabular}
\footnotetext[1]{
Effective momentum transfer to the $\Lambda$ final state, 
$q_{\rm eff}\approx (M_{\rm C}/M_{\rm A})q$ given in Eq.~(\ref{eqn:e25}).
}
\end{table*}

The energy and momentum transfer to the $\Lambda$ final state is given by
\begin{eqnarray}
\omega = E_\pi- E_K, \qquad  {\bm q} ={\bm p}_{\pi}-{\bm p}_{K},
\label{eqn:e4}
\end{eqnarray}
where 
$E_\pi=(m_\pi^2+{\bm p}_\pi^2)^{1/2}$ and 
$E_K=(m_K^2+{\bm p}_K^2)^{1/2}$ are the lab energies 
of $\pi$ and $K$ in the nuclear reaction, respectively; $m_\pi$ and $m_K$ 
(${\bm p}_{\pi}$ and ${\bm p}_{K}$) are
masses (lab momenta) of $\pi$ and $K$, respectively. 
Figure \ref{fig:1} displays the momentum transfer 
to the $0^+_{\rm g.s.}$ final state in $^4_\Lambda$He, 
\begin{eqnarray}
q =(p_\pi^2+p_K^2-2p_\pi p_K \cos{\theta_{\rm lab}})^{1/2},
\label{eqn:e5}
\end{eqnarray}
as a function of the incident lab momentum $p_\pi$.  

The differential lab cross section of a $\Lambda$ hypernuclear 
state is obtained by the energy integration \cite{Morimatsu94} 
\begin{eqnarray}
\biggl({d\sigma \over d{\Omega}_K}\biggr)
&=& \int {dE_K}\, 
\biggl({d^2\sigma \over dE_K d{\Omega}_K}\biggr),  
\label{eqn:e6}
\end{eqnarray}
around a corresponding peak in the inclusive $K$ spectrum.
We often adopt the effective number technique into 
the differential lab cross section within DWIA 
\cite{Hufner74,Dover83,Koike08,Morimatsu94,Itonaga94}. 
Thus the differential lab cross section of the $\Lambda$ bound state with 
$J^\pi$ can be written as 
\begin{eqnarray}
&&\left({d\sigma \over d\Omega_K}\right)_{\rm lab, \theta_{\rm lab}}^{J^\pi}
=  \alpha {1 \over {[J_A]}} \sum_{m_Am_B}
\biggl| \Bigl\langle {\Psi}_B 
\Big\vert\, \overline{f}_{\pi N \to K \Lambda}        \nonumber\\
&& \qquad \times \,
\chi^{(-)*}_{K}({\bm p}_{K},{M_{C} \over M_{B}}{\bm r}) 
\chi^{(+)}_{\pi}({\bm p}_{\pi},{M_{C} \over M_{A}}{\bm r}) 
\Big| \Psi_{A} \Bigr\rangle \biggr|^2, \qquad
\label{eqn:e7}
\end{eqnarray}
where 
${\bm r}$ is the relative coordinate between a $3N$-core nucleus 
and a nucleon or $\Lambda$ hyperon; 
the factors of $M_{C}/M_{B}$ and $M_{C}/M_{A}$ 
take into account the recoil effects where $M_{A}$, $M_{B}$, and $M_{C}$ are masses
of the target, the hypernucleus, and the core nucleus, respectively.
The kinematical factor $\alpha$ \cite{Dover83} is related to $\beta$ in Eq.~(\ref{eqn:e2}) as
\begin{equation}
\alpha= \beta \left(1+{E_K \over E_B}{p_K-p_\pi 
\cos{\theta_{\rm lab}} \over p_K}\right)^{-1}.
\label{eqn:e8}
\end{equation}

In Table~\ref{tab:1}, we show the kinematic values for 
$^4_\Lambda$He production via the nuclear ($\pi$,~$K$) 
reaction on a $^4$He target at $p_\pi=$ 1.05 GeV/$c$, 
$\theta_{\rm lab}=$ 0$^\circ$, 4$^\circ$, 8$^\circ$, and 12$^\circ$,
where we choose $B_\Lambda=$ 2.39 MeV as the $\Lambda$ binding energy 
for $0^+_{\rm g.s}$ in $^4_\Lambda$He. 
Note that the value of $\alpha$ for a $^4$He target 
at $p_\pi=$ 1.05 GeV/$c$ (0$^\circ$) amounts to 0.828 which is 25\%
larger than that for heavier nuclei; $\alpha\simeq$ 0.66 
for a $^{40}$Ca target \cite{Dover80}.

\subsection{Fermi-averaged ($\pi$,~$K$) cross sections}
\label{sect:2-2}

It should be noticed that the strong energy dependence of differential lab cross sections 
appears in the nuclear ($\pi$,~$K$) reactions, as discussed in Ref.~\cite{Harada05}. 
To consider the $p_\pi$ dependence of elementary $\pi N\to K \Lambda$ lab
cross sections, we do averaging of the $\pi N\to K\Lambda$ $t$ matrix 
in the lab frame over a Fermi-momentum distribution, where nuclear effects of 
a nucleon binding $\varepsilon_N$ are naturally taken into account.
This procedure is called as the ``optimal Fermi-averaging'' under 
the on-energy-shell condition \cite{Harada05}. 
Charge independence guarantees the following relation between 
the $\pi N\to K\Lambda$ amplitudes:
\begin{equation}
{f}_{\pi^+ n\to K^+\Lambda}=-{f}_{\pi^- p\to K^0\Lambda}.
\label{eqn:e9}
\end{equation}
Thus the ($\pi^+$,~$K^+$) and ($\pi^-$,~$K^0$) cross sections are identical to each other 
on the $T=0$ nuclear targets as $^4$He. 
Here we employed the elementary $\pi^- p\to K^0\Lambda$ amplitudes analyzed by Sotona 
and $\check{\rm Z}$ofka \cite{Sotona89}. 

Figure~\ref{fig:2} shows the optimal Fermi-averaged $\pi N\to K\Lambda$ lab cross sections of 
$\alpha \langle d\sigma/d\Omega \rangle^{\rm opt}_{\pi N \to K \Lambda}$ 
including the kinematical factor $\alpha$ in nuclear ($\pi$,~$K$) reactions on a $^{4}$He target, 
together with the elementary $\pi N\to K\Lambda$ lab cross sections 
of $\alpha \langle d\sigma/d\Omega \rangle^{\rm free}_{\pi N \to K \Lambda}$ 
in free space \cite{Sotona89}. 
The peaks of $\alpha \langle d\sigma/d\Omega \rangle^{\rm free}_{\pi N \to K \Lambda}$
are located at $p_\pi \simeq$ 1.05 GeV/$c$ corresponding to 
$M(\pi N)\simeq$ 1700 MeV/$c^2$ in the invariant mass of $\pi$ and $N$
because there exist $N^*$ resonances, e.g., $S_{11}$(1680), $P_{11}$(1730), and $P_{13}$(1700). 
We find that the peaks of 
$\alpha \langle d\sigma/d\Omega \rangle^{\rm opt}_{\pi N \to K \Lambda}$ 
are shifted to the position of $p_\pi \simeq$ 1.00 GeV/$c$, 
taken into account the Fermi motion of a struck nucleon 
under the optimal condition in the nucleus; 
the shape of $\alpha \langle d\sigma/d\Omega \rangle^{\rm opt}_{\pi N \to K \Lambda}$ 
is moderately broader than that of 
$\alpha \langle d\sigma/d\Omega \rangle^{\rm free}_{\pi N \to K \Lambda}$. 
In Table~\ref{tab:1}, we also show the values of 
$\alpha \langle d\sigma/d\Omega \rangle_{\pi N\to\Lambda K}^{\rm opt}$, together with 
the kinematical factor $\alpha$ at $p_\pi =$ 1.05 GeV/$c$.

\subsection{Wavefunctions}
\label{sect:2-3}

\begin{figure}[b]
\begin{center}
  \includegraphics[width=0.8\linewidth]{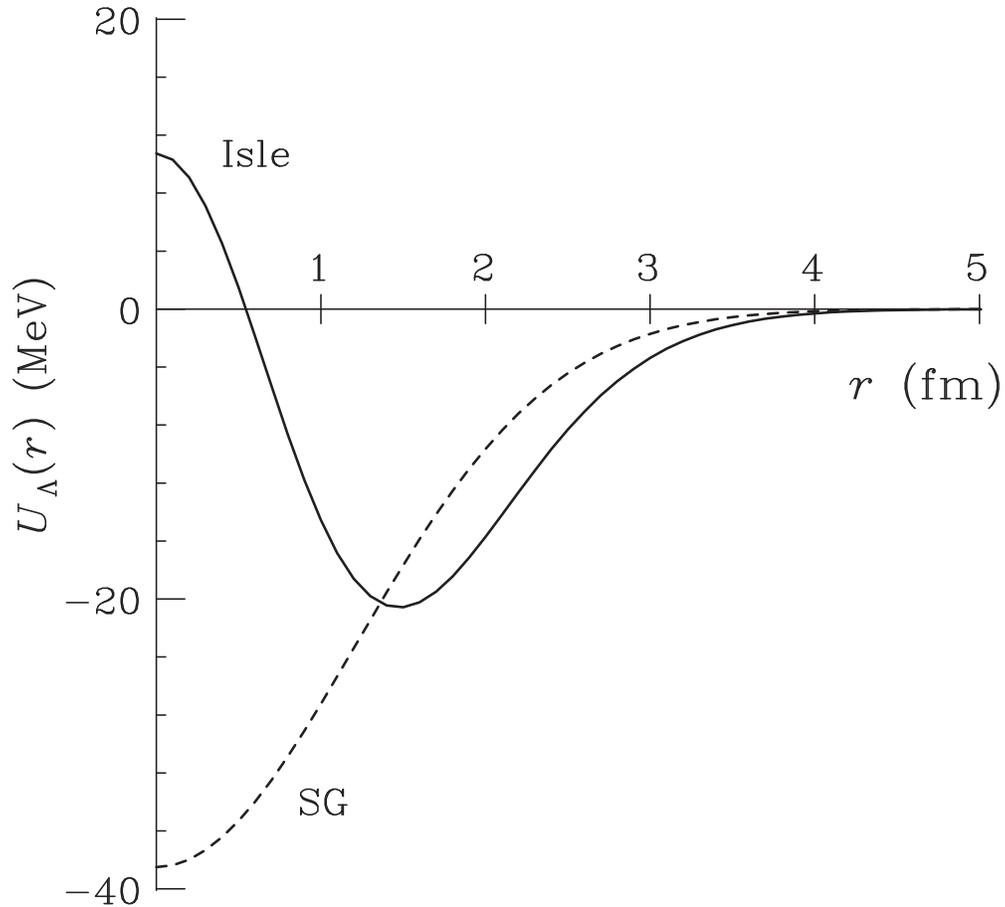}
\end{center}
\caption{\label{fig:3}
The $3N$-$\Lambda$ potential $U_\Lambda$ for 0$^+_{\rm g.s.}$ in $^4_\Lambda$He, 
as a function of the relative distance between $3N$ and $\Lambda$.
Solid and dash curves denote the Isle and SG potentials, respectively.  
}
\end{figure}

In our calculations, we have assumed the $3N$ $1/2^+$ ground state as a core-nucleus 
in $A=$ 4 systems \cite{Harada98}, 
i.e., $\phi_{3N}$ is a wavefunction for the $3N$ 
eigenstates ($^3$He or $^3$H), and we have neglected the CSB effects in $A=$ 4 hypernuclei.
Thus the wavefunction of the $0^+$ ground state ($0^+_{\rm g.s.}$)
in $^4_\Lambda$He ($L_B=$ 0, $S_B=$ 0, $T_B=$ 1/2)
is written as 
\begin{eqnarray}
&&{\Psi}_B  
= \left[\bigl[\phi_{3N}\otimes\varphi^{(\Lambda)}_{\ell_\Lambda}\bigr]_{L_B}
  \otimes X^B_{T_B,S_B}\right]_{J_B}, \label{eqn:e12}\\
&&X^B_{T_B,S_B}=
\bigl[\chi^{(3N)}_{I_3,S_3}\otimes \chi^{(\Lambda)}_{{0},{1/2}}\bigr]_{1/2,0},  \nonumber
\end{eqnarray}
where $\varphi^{(\Lambda)}_{\ell_\Lambda}$ is a relative wavefunction 
between $3N$ and $\Lambda$, 
and $X^B_{T_B,S_B}$ is the isospin-spin function for $0^+_{\rm g.s.}$ in $^4_\Lambda$He; 
$\chi^{(3N)}_{I_3,S_3}$ and $\chi^{(\Lambda)}_{{0},{1/2}}$
are the isospin-spin functions for $3N$ (isospin $I_3$, spin $S_3$) and 
$\Lambda$ ($I=$ 0, $S=$ 1/2), respectively.
The explicit form of $X^B_{T_B,S_B}$ is given in Appendix \ref{app:1}.
According to Ref.~\cite{Kurihara82}, we use $\varphi^{(\Lambda)}_{\ell_\Lambda}$ 
which is regarded as a spectroscopic amplitude obtained from a four-body $\Lambda NNN$ wavefunction 
using realistic central nucleon-nucleon ($NN$) and $\Lambda N$ potentials \cite{Dalitz72}.
It should be noticed that $\varphi^{(\Lambda)}_{\ell_\Lambda}$ includes the contribution of 
the $\Lambda N$ short-range correlations and also many-body correlations.
Thus we consider that $\varphi^{(\Lambda)}_{\ell_\Lambda}$ satisfies the Schr\"odinger equation
\begin{eqnarray}
&& \left(-{\hbar^2 \over 2\mu}\nabla^2 + U_\Lambda \right) \varphi^{(\Lambda)}_{\ell_\Lambda} 
= -B_\Lambda \varphi^{(\Lambda)}_{\ell_\Lambda},
\label{eqn:e1}
\end{eqnarray}
where $\mu$ is the $3N$-$\Lambda$ reduced mass and $B_\Lambda$ is the $\Lambda$ binding 
energy with respect to the $^3{\rm He}$-$\Lambda$ threshold. 
The $3N$-$\Lambda$ potential ${U}_{\Lambda}$ is defined as 
\begin{eqnarray}
{U}_{\Lambda}
&=& \langle \phi_{3N} | \bar{V}^{\rm ex}\hat{F}^{\rm ex} | \phi_{3N} \rangle, 
\label{eqn:e11a}
\end{eqnarray}
where 
$\bar{V}^{\rm ex}$ is the sum of isospin-spin averaged $\Lambda N$ potentials, 
and $\hat{F}^{\rm ex}$ is an external operator on the basis of multiple scattering processes \cite{Akaishi86}:
\begin{eqnarray}
\hat{F}^{\rm ex} &=& 1+ \frac{Q}{e}\bar{V}^{\rm ex}\hat{F}^{\rm ex}, 
\label{eqn:e11b}
\end{eqnarray}
where $Q=1-|\phi_{3N})(\phi_{3N}|$ and $e=E-\hat{H}_{3N}+\hbar^2\nabla^2/2\mu$. 
Therefore, ${U}_{\Lambda}$ can be derived from the four-body $\Lambda NNN$ wavefunction,
taken into account the $\Lambda N$ short-range correlations \cite{Kurihara82,Akaishi86}. 

Figure~\ref{fig:3} shows ${U}_{\Lambda}$ as a function of the distance
between $3N$ and $\Lambda$. 
We recognize that this potential has a central repulsion 
and an attractive tail, so we call it as ``Isle'' potential \cite{Fuse95,Kurihara82,Akaishi86,Kurihara85}; 
the central repulsion originates predominantly from the $\Lambda N$ short-range correlations 
due to the repulsive core of the $\Lambda N$ potentials, 
and it plays an important role in describing 
the lifetime of $^4_\Lambda$He in precise experimental studies 
on the mesonic weak decay of $\Lambda \to p\pi^-$ \cite{Outa98,Fuse95}.
For convenience of use, we parameterize $U_\Lambda$ into a two-range Gaussian form as 
\begin{eqnarray}
U_{\Lambda}({\bm r}) 
&=& V_{C}\exp \{ -(r/b_{C})^2 \} \nonumber\\
&& \quad +V_{A}\exp \{-({r/b_{A}})^2 \}, 
\label{eqn:e13}
\end{eqnarray}
where $V_{C}=$ 91.61 MeV, $V_{A}=$ $-$80.88 MeV, 
$b_{C}=$ 1.14 fm, and $b_{A}=$ 1.69 fm, 
reproducing the experimental data of $B_\Lambda^{\rm exp}=$ $2.39\pm0.03$ MeV.
To clearly see the effect of the central repulsion, 
we also introduce a single-range Gaussian 
(SG) potential of $U_\Lambda({\bm r})=-38.5 \exp \{-(r/b_A)^2 \}$
with $b_A=$ 1.70 fm, adjusting to $B_\Lambda=$ 2.39 MeV.  
The SG potential is often used as a phenomenological one. 

The wavefunction of the $0^+$ ground state in $^{4}$He 
($L_A=$ 0, $S_A=$ 0, $T_A=$ 0) is written as 
\begin{eqnarray}
&&{\Psi}_A  
= {\cal A}\left[\bigl[\phi_{3N}\otimes\varphi^{(N)}_{\ell_N}\bigr]_{L_A}
    \otimes X^A_{T_A,S_A}\right]_{J_A}, 
\label{eqn:e10}\\
&& X^A_{T_A,S_A}=
\bigl[\chi^{(3N)}_{I_3,S_3}\otimes \chi^{(N)}_{{1/2},{1/2}}\bigr]_{0,0}, \nonumber
\end{eqnarray}
where ${\cal A}$ is the anti-symmetrized operator for nucleons, and 
$\varphi^{(N)}_{\ell_N}$ is a relative wavefunction between $3N$ and $N$;
$X^A_{T_A,S_A}$ and $\chi^{(N)}_{{1/2},{1/2}}$ are the isospin-spin functions 
for $^4$He and $N$ ($I =$ 1/2, $S =$ 1/2), respectively. 
Here we used $\varphi^{(N)}_{\ell_N}$ obtained from the $3N$-$N$ potential ${U}_{N}$ 
which was derived from a microscopic four-body calculation 
with a central $NN$ potential of Tamagaki's C3G \cite{Akaishi86}.
Therefore, $\varphi^{(N)}_{\ell_N}$ includes the contribution of 
the $NN$ short-range correlations and also many-body correlations \cite{Kurihara82}. 
For convenience of use, this potential ${U}_{N}$ is parametrized into useful Gaussian forms as 
\begin{eqnarray}
{U}_{N}({\bm r}) 
&=& V_{1} \exp \{-(r/b_{1})^2 \}
+ V_{2} \exp \{-(r/b_{2})^2 \} \nonumber\\
&& \qquad + V_{3} \exp \{-(r/b_{3})^2 \}
\label{eqn:e11}
\end{eqnarray}
with $V_{1}=$ 156.28 MeV, $V_{2}=$ $-$185.66 MeV, $V_{3}=$ $-$9.56 MeV,
$b_{1}=$ 1.21 fm, $b_{2}=$ 1.58 fm, and $b_{3}=$ 2.82 fm \cite{Harada90}, 
making a fit to the experimental data of the binding energy 
$B_{N}=$ 20.6 MeV and the nuclear root-mean-square (r.m.s.) distance 
of $\langle {r}^2_N \rangle^{1/2}=$ 1.87 fm 
between $^3{\rm He}$ and $n$. 

\begin{figure}[t]
\begin{center}
  \includegraphics[width=\linewidth]{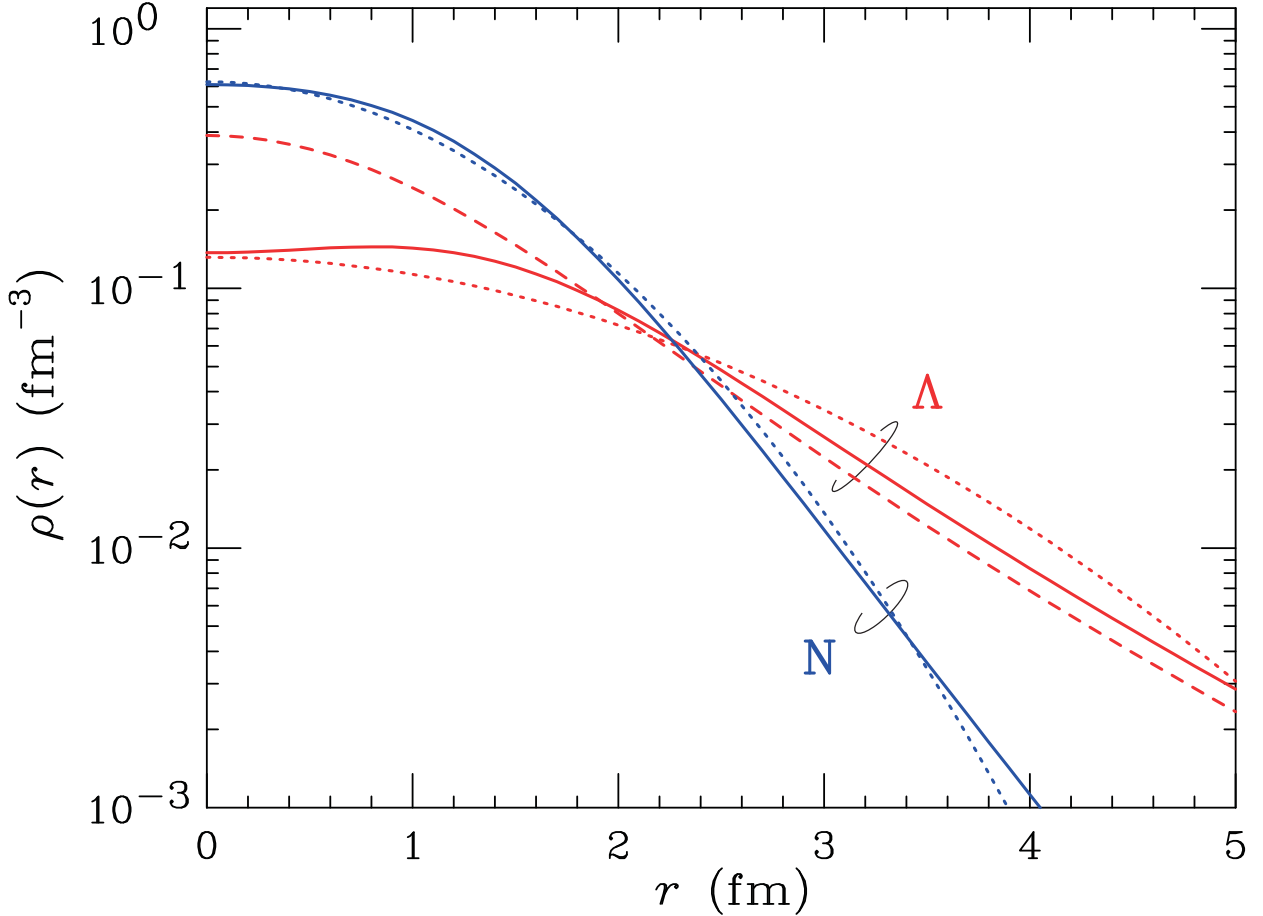}
\end{center}
\caption{\label{fig:4}
$3N$-$\Lambda$ density distributions $\rho_\Lambda(r)$ for $0^+_{\rm g.s.}$ in $^4_\Lambda$He, 
as a function of the relative distance, 
together with $3N$-$N$ density distributions $\rho_N (r)$ for $0^+_{\rm g.s.}$ in $^4$He. 
Solid and dashed curves for $\Lambda$ denote the distributions for 
the Isle and SG potentials, respectively. 
The solid curve for $N$ denotes the distribution for the C3G potential. 
Dotted curves are the distributions for HO with the c.m.~corrections. 
}
\end{figure}

Figure~\ref{fig:4} shows that $3N$-$\Lambda$ density distributions $\rho_\Lambda(r)$ 
between $3N$ and $\Lambda$ for $0^+_{\rm g.s.}$ in $^4_\Lambda$He, 
as a function of the relative distance, together with $3N$-$N$ density distributions 
$\rho_N(r)$ between $3N$ and $N$ for $0^+_{\rm g.s.}$ in $^4$He.
We find that the $3N$-$\Lambda$ distribution obtained from the Isle potential
differs fairly from that obtained from the SG potential around the nuclear inside;
the $3N$-$\Lambda$ distribution is considerably suppressed at the nuclear center, 
and it is pushed outside, as discussed in Refs.~\cite{Kurihara82,Akaishi86,Kurihara85,Harada90}.
The relative r.m.s.~distance between $3N$ and $\Lambda$ becomes 
$\langle {r}^2_\Lambda \rangle^{1/2}=$ 3.57 fm for Isle, which is 8\% larger than 3.31 fm for SG. 

To see properties of our $\varphi^{(\Lambda)}_{\ell_\Lambda}$ obtained by $U_\Lambda$ in Fig.~\ref{fig:3}, 
we also consider a single-particle (s.p.)~harmonic oscillator (HO) wavefunction
with the size parameter $b_\Lambda=(\hbar/m_\Lambda\omega_\Lambda)^{1/2}$, 
which is used as simple model calculations \cite{Dover80,Cieply01}.
Here we choose $b_\Lambda=$ 2.233 fm ($\hbar \omega_\Lambda=$ 7.0 MeV), 
simulating the value of the relative r.m.s.~distance 
of $\langle r^2_\Lambda \rangle^{1/2}=$ 3.3--3.6 fm 
for the $3N$-$\Lambda$ distribution obtained by four-body $\Lambda NNN$ calculations 
using the $NN$ and $\Lambda N$ potentials \cite{Dalitz72,Bodmer85,Hiyama01}; 
we have $\langle r^2_\Lambda \rangle^{1/2}=$ 3.23 fm for HO.
The relative $3N$-$\Lambda$ wavefunction with ${\ell_\Lambda}=0$ 
for HO is written as 
\begin{eqnarray}
\varphi^{(\Lambda)}_{\ell_\Lambda}({\bm r})=
{\biggl(\frac{4}{\tilde{b}_\Lambda^3\sqrt{\pi}}\biggr)^{1/2}} 
\exp{\biggl(-\frac{r^2}{2\tilde{b}_\Lambda^2}\biggr)}Y_{0}^{0}(\hat{\bm r}),
\label{eqn:e13b}
\end{eqnarray}
where the size parameter with the c.m.~correction denotes
\begin{eqnarray}
\tilde{b}_\Lambda=b_\Lambda\sqrt{1+m_\Lambda/3m_N}=2.634 \,{\rm fm},
\label{eqn:e13a}
\end{eqnarray}
where $m_\Lambda$ and $m_N$ are masses of a $\Lambda$ hyperon and a nucleon, respectively.
We confirm $\langle r^2_\Lambda \rangle^{1/2}=\tilde{b}_\Lambda\sqrt{3/2}=$ 3.23 fm for HO.
Figure~\ref{fig:4} also shows the $3N$-$\Lambda$ distribution 
for HO taken into account 
the c.m.~correction which is absolutely essential in light nuclei. 
We expect that the difference among the $3N$-$\Lambda$ distributions obtained from several models 
in Fig.~\ref{fig:4} is clearly observed in production cross sections of nuclear ($\pi$, $K$) reactions
giving a large momentum transfer to the $\Lambda$ final state.
On the other hand, we confirm that the $3N$-$N$ distribution for HO 
using the size parameter $b_N=$ 1.329 fm ($\hbar \omega_\Lambda=$ 23.5 MeV)
is in good agreement with that obtained from the C3G potential; 
the relative r.m.s.~distance between $3N$ and $N$ has 
$\langle r^2_N \rangle^{1/2}=\tilde{b}_N\sqrt{3/2}=$ 1.88 fm 
which is quite close to 1.87 fm for C3G, where $\tilde{b}_N=b_N \sqrt{4/3}=$ 1.535 fm.

\subsection{Differential cross sections}
\label{sect:2-4}

We consider the differential lab cross sections of  
the hypernuclear bound state with $J^\pi$ on 
a closed-shell target nucleus with $J_A^\pi=$ 0$^+_{\rm g.s.}$ like $^4$He, 
adapting the effective number technique into the 
DWIA \cite{Hufner74,Dover83,Koike08,Morimatsu94,Itonaga94}.
Substituting Eqs.~(\ref{eqn:e10}), (\ref{eqn:e12}), 
and (\ref{eqn:e22}) into Eq.~(\ref{eqn:e7}), we obtain 
\begin{equation}
\left({d\sigma \over d\Omega_K}\right)^{J^\pi}_{{\rm lab},\theta_{\rm lab}}
= \alpha \left\langle d\sigma \over d\Omega_K \right\rangle^{\rm opt}_{\pi N \to K\Lambda}
N^{J^\pi}_{\rm eff}(\theta_{\rm lab}),
\label{eqn:e14}
\end{equation}
where $\alpha\langle d\sigma/d\Omega_K \rangle_{\pi N \to K\Lambda}^{\rm opt}$ is 
the optimal Fermi-averaged $\pi N \to K \Lambda$ lab cross section, 
as discussed in Sect.~\ref{sect:2-2}. 
The effective number of nucleons $N^{J^\pi}_{\rm eff}$ 
for $\Lambda$ production of the $J^\pi$ final state in the $LS$-coupling scheme 
is written as
\begin{eqnarray}
    N^{J^\pi}_{\rm eff}(\theta_{\rm lab})
&=& C^2_{B}(2L+1)(2\ell_\Lambda+1) \nonumber\\
&& \times
    \left( \begin{array}{ccc}
     \ell_\Lambda & L & \ell_N  \\
     0 & 0 & 0 
    \end{array} \right)^2
    |F(q)|^2, 
\label{eqn:e15}
\end{eqnarray}
where $\ell_\Lambda+L+\ell_N$ must 
be even due to the non-spin-flip reaction. 
Thus, only natural parity states with 
$J^\pi=$ $0^+$, $1^-$, $2^+$, $3^-$, 
$\cdots$ for the $3N+\Lambda$ systems can be populated.
$C_B$ is the isospin-spin spectroscopic amplitude between 
the $\Lambda$ final state of $^4_\Lambda$He 
and the initial state of $^4{\rm He}$, which is given by
\begin{equation}
C_{B}= \Bigl\langle 
        X^B_{T_B,S_B} \Big| 
      \sum_{j=1}^{A}{\hat O}_{j}  
      \Big|
        X^A_{T_A,S_A}
      \Bigr\rangle.
\label{eqn:e16} 
\end{equation}
The form factor $F(q)$ in Eq.~(\ref{eqn:e15}) is given as 
\begin{eqnarray}
F(q)&=&
\int_0^{\infty} r^2d{r} 
\rho^{(tr)}_{\ell_\Lambda\ell_N}(r)\widetilde{j}_{L}(q;\frac{M_C}{M_A}r)
\label{eqn:e17} 
\end{eqnarray}
with the $N\to\Lambda$ transition densities
\begin{eqnarray}
\rho^{(tr)}_{\ell_\Lambda\ell_N}(r)
&=&{\varphi}^{(\Lambda)*}_{\ell_{\Lambda}}(r)\varphi^{(N)}_{\ell_N}(r)
\label{eqn:e18}
\end{eqnarray}
and the distorted waves $\widetilde{j}_{L}(q; z)$ 
considering the nuclear distortions by mesons in the DW approximation, 
as we will express in Eq.~(\ref{eqn:e23}); 
${M_C/M_A}$ is the so-called recoil factor for the momentum transfer $q$. 
Here we approximated to $M_C/M_A \approx M_C/M_B$ in Eq.~(\ref{eqn:e7}). 
For $J^\pi=$ $0^+_{\rm g.s.}$ in $^4_\Lambda$He, 
we take $\ell_\Lambda=\ell_N=L=0$ and $C_B=\sqrt{2}$ in Eq.~(\ref{eqn:e15}), 
then we obtain
\begin{equation}
N^{0^+}_{\rm eff}\!(\theta_{\rm lab})=2|F(q)|^2, 
\label{eqn:e19}
\end{equation}
which is expected to observe the hypernuclear fine structure 
because $F(q)$ is generally sensitive to 
the nature of the distribution of $\rho^{(tr)}_{00}(r)$, 
as a function of $q$.

\subsection{Meson distorted waves}
\label{sect:2-5}

\begin{figure}[tb]
\begin{center}
  \includegraphics[width=\linewidth]{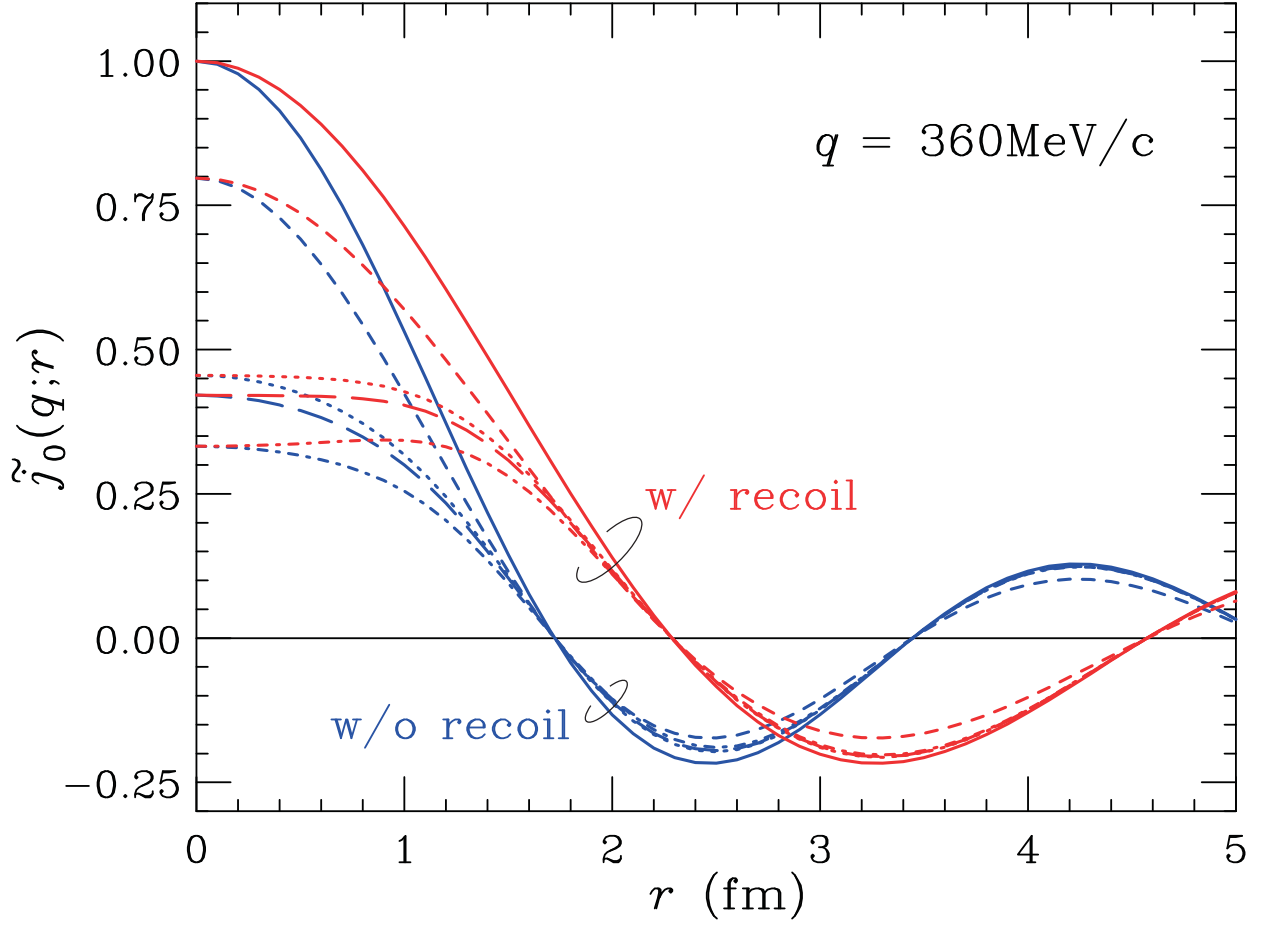}
\end{center}
\caption{\label{fig:5}
Distorted waves $\widetilde{j}_L(q;r)$ with $L=0$ for $\pi$ and $K$ 
in the $^4$He($\pi$,~$K$) reaction at 
$p_{\pi}=$ 1.05 GeV/$c$ ($\theta_{\rm lab}=$ 0$^\circ$) which leads to $q=$ 360 MeV/$c$. 
Solid and dashed curves denote the distorted and the plane waves with/without recoil effects, respectively, 
as a function of the radial distance between the mesons and the center of the nucleus.
Dotted curves denote the waves in the eikonal-oscillator approximation with $\bar{\sigma}=$ 30 mb.  
}
\end{figure}

Full distorted waves of the $\pi$-nucleus and the $K$-nucleus 
are important to reproduce absolute values of the cross sections.
Because the ($\pi$,~$K$) reaction requires a large momentum transfer 
with a high angular momentum, 
we simplify the computational procedure in the eikonal approximation 
to the distorted waves of the meson-nucleus 
states \cite{Hufner74,Dover80,Dover83,Koike08}: 
\begin{equation}
\chi^{(-)*}_{K}({\bm p}_K, {\bm r})\chi^{(+)}_{\pi}({\bm p}_\pi, {\bm r}) 
=  \exp{(i{\bm q}\cdot{\bm r})}D({\bm b},z)
\label{eqn:e20}
\end{equation}
with
\begin{eqnarray}
D({\bm b},z) 
&=& \exp{\biggl( 
  -{\sigma_\pi(1-i \alpha_\pi) \over 2} \int^z_{-\infty} \rho({\bm b},z')dz'} \nonumber\\
&& { -{\sigma_K(1+i \alpha_K) \over 2} \int_z^{\infty} \rho({\bm b},z')dz'\biggr)},
\label{eqn:e21}
\end{eqnarray}
where $\sigma_\pi$ ($\sigma_K$) is the averaged total cross section in $\pi N$ ($K N$) elastic scatterings, 
and $\alpha_\pi$ ($\alpha_K$) is 
the ratio of the real and imaginary part of the corresponding 
forward scattering amplitudes; ${\bm b}$ is the impact parameter. 
$\rho({\bm r})\equiv\rho({\bm b},z)$ is a matter-density distribution
fitting to the data on the nuclear charge density \cite{Vries86}. 
We assume $\alpha_\pi = \alpha_K =$ 0 which affect hardly the following results, 
and we study $\sigma_\pi$= 20--30 mb and $\sigma_K$= 10--30 mb~\cite{Hufner74,Dover80}; 
($\sigma_\pi$, $\sigma_K$)= (30 mb, 15 mb) is chosen as a standard value \cite{Harada05}.
Reducing the r.h.s.~in Eq.~(\ref{eqn:e20}) by partial-wave expansion, 
we obtain  
\begin{eqnarray}
\chi^{(-)*}_{K}({\bm p}_K, {\bm r})\chi^{(+)}_{\pi}({\bm p}_\pi, {\bm r}) 
&=&\sum_L \sqrt{4 \pi (2L+1)} i^L \nonumber\\
&& \times \widetilde{j}_L(q;r)Y_L^0(\hat{\bm r}) 
\label{eqn:e22}
\end{eqnarray}
with 
\begin{eqnarray}
\widetilde{j}_L(q;r)
&=&\sum_{\ell \ell'}i^{\ell-L} {2\ell +1 \over 2L+1}\sqrt{2\ell'+1}
(\ell 0 \ell' 0|L0)^2  \nonumber\\
&& \times j_{\ell}(qr)D_{\ell'}(r),
\label{eqn:e23}
\end{eqnarray}
where  $D_\ell(r)$ is a distortion function \cite{Harada05} defined as 
\begin{equation}
D_{\ell}(r) 
= {{2 \ell +1} \over 2}
\int^{1}_{-1} D({\bm b},z) P_\ell(\cos{\theta}) d(\cos{\theta}).
\label{eqn:e23a}
\end{equation}
Here $z= r \cos{\theta}$, ${\bm r}^2={\bm b}^2+z^2$, and $P_\ell(x)$ 
is a Legendre polynomial. 
If the distortion is switched off, 
$\widetilde{j}_{L}(q;r)$ is equal to ${j}_{L}(qr)$ 
which is a spherical Bessel function with $L$.


The production probability for $0^+_{\rm g.s.}$ in $^4_\Lambda$He is expected to be only
$10^{-2}$, which is roughly estimated as $\exp{(-(bq)^2/2)}$ with $bq \simeq 3$ \cite{Dover80}, 
due to the $0s_N \to 0s_\Lambda$ transition with $\Delta L=$ 0 in nuclear ($\pi$, $K$) reactions
because the $\Lambda$ continuum states can be populated predominately by the large angular momentum transfer. 
Figure~\ref{fig:5} displays the distorted waves $\widetilde{j_0}(q;r)$ for $\pi$ and $K$ 
in the $^4$He($\pi$,~$K$) reactions at $p_{\pi}=$ 1.05 GeV/$c$ ($\theta_{\rm lab}=$ 0$^\circ$) 
which leads to $q=$ 360 MeV/$c$.
We find that the values of $\widetilde{j_0}(q;r)$ are reduced 
near the center of the nucleus due to the nuclear absorption in the distorted wave, 
in comparison with the plane waves which are obtained with  
($\sigma_\pi$, $\sigma_K$) = (0 mb, 0 mb).
We also find that $\widetilde{j_0}(q;r)$ spread outside
by taking into account the recoil effects 
which bring us to use the {\it effective} momentum transfer
\begin{eqnarray}
q_{\rm eff}
&=& \frac{M_C}{M_A}q  \nonumber\\
&\simeq& \frac{A-1}{A}q = \frac{3}{4}\times 360 = 271 \ \mbox{MeV/c}
\label{eqn:e25}
\end{eqnarray}
in the $A=4$ hypernuclei, as seen in Table~\ref{tab:1}. 
We recognize a node in $\widetilde{j_0}(q;r)$ at $r=r_n$ 
satisfied as 
\begin{eqnarray}
\frac{M_C}{M_A}q r_n = n\pi, \qquad (n= 1, 2, \cdots),
\label{eqn:e24}
\end{eqnarray}
so we have the point of $r_1 =(M_A/M_C)\pi/q=$ 2.29 fm located on the outside of 
the nucleus having $\langle r^2_N \rangle^{1/2}=$ 1.88 fm for $^4$He.  
If we omit the recoil effects ($M_C/M_A \to 1$), 
we have $r_1 \to \pi/q =$ 1.72 fm which is located on the 
inside of the nucleus. 
Therefore, the recoil effects must be taken into account 
for the light nuclear target in the ($\pi$,~$K$) reactions 
providing the large momentum transfer to the $\Lambda$ final state,
as we will discuss in Sect.~\ref{recoil}.

\begin{table}[t]
\caption{\label{tab:2}
Calculated PWIA and DWIA results of the differential lab cross sections of $0^+_{\rm g.s.}$ 
in $^4_\Lambda$He via the $^4$He($\pi^+$,~$K^+$) reaction at $p_{\pi}=$ 1.05 GeV/$c$.
The Isle, SG, and HO ($\hbar\omega_\Lambda=$ 7.0 MeV) potentials are used with 
$({\sigma}_\pi,~{\sigma}_K)=$ (30 mb,~15 mb).
The recoil effects are taken into account.
}
\begin{ruledtabular}
\begin{tabular}{lccccccc}
\noalign{\smallskip}
& $\theta_{\rm lab}$
& {$N_{\rm eff}^{\rm PW}$}
& {$(d\sigma/d\Omega)_{\rm lab}^{\rm PW}$}
& {$D_{\rm dis}$\footnotemark[1]}
& {$N_{\rm eff}^{\rm DW}$ }      
& {$(d\sigma/d\Omega)_{\rm lab}^{\rm DW}$} \\
& (degree)
& {($\times 10^{-2}$)}
& {($\mu$b/sr)} 
& 
& {($\times 10^{-2}$)} 
& {($\mu$b/sr)}   \\
\noalign{\smallskip}\hline\noalign{\smallskip}
\multicolumn{3}{l}{Isle}   \\
&0 & 7.571   & 35.14   & 0.334   & 2.527   & 11.73   \\
&4 & 6.951   & 33.88   & 0.324   & 2.254   & 10.99   \\
&8 & 5.390   & 25.25   & 0.296   & 1.595   & \ \ 7.47   \\
&12 & 3.540  & 14.72   & 0.248   & 0.879   & \ \ 3.65   \\
\noalign{\smallskip}
\multicolumn{3}{l}{SG}   \\
&0 & 12.68   & 58.84   & 0.332   & 4.211   &  19.55   \\
&4 & 11.82   & 57.62   & 0.325   & 3.841   &  18.72   \\
&8 & 9.616   & 45.06   & 0.303   & 2.918   &  13.67   \\
&12 & 6.876  & 28.58   & 0.268   & 1.846   &  \ \ 7.67   \\
\noalign{\smallskip}
\multicolumn{5}{l}{HO}   \\
& 0  & 5.160  &	23.95 &  0.309  & 1.595 & \ \ 7.40 \\
& 4  & 4.704  &	22.93 &  0.298  & 1.402 & \ \ 6.83 \\
& 8  & 3.574  &	16.75 &  0.265  & 0.948 & \ \ 4.44 \\
& 12  & 2.281 &	9.481 &  0.211  & 0.482 & \ \ 2.00 \\
\noalign{\smallskip}
\end{tabular}
\end{ruledtabular}
\footnotetext[1]{
$D_{\rm dis}= N^{\rm DW}_{\rm eff}/N^{\rm PW}_{\rm eff}$.}
\end{table}

\section{Results and Discussion}

Let us consider the $\Lambda$ production for $0^+_{\rm g.s.}$ in $^4_\Lambda$He 
via the ($\pi$,~$K$) reactions on the $^4$He target
at $p_\pi=$ 1.05 GeV/$c$ in the $K$ forward direction.
We will discuss the meson distortion effects, 
comparing between the differential lab cross sections in PWIA and DWIA, 
and we will study the sensitivity of the cross sections to 
the $\Lambda$ wavefunctions using the effective number technique of Eq.~(\ref{eqn:e14}). 

\subsection{Differential cross sections for $0^+_{\rm g.s.}$ in $^4_\Lambda$He}

\subsubsection{PWIA v.s. DWIA}

In Table~\ref{tab:2}, 
we show the calculated PWIA and DWIA results of the differential lab cross 
sections for $0^+_{\rm g.s.}$ in $^4_\Lambda$He at $p_\pi=$ 1.05 GeV/$c$ 
in the $K$ forward-direction angles of $\theta_{\rm lab}=$ 0$^\circ$--12$^\circ$. 
The differential lab cross section of $0^+_{\rm g.s.}$ in $^4_\Lambda$He
accounts for 
\begin{eqnarray}
  \left({d\sigma \over d\Omega_K}\right)^{J^\pi=0^+}_{\rm lab, 0^\circ}
  &=& 35.14 \,\mbox{$\mu$b/sr} \qquad \mbox{(PWIA)}
\label{eqn:e32}
\end{eqnarray}
at $p_\pi=$ 1.05 GeV/$c$ ($\theta_{\rm lab}=$ $0^\circ$), using the Isle potential. 
When we take into account the distortion with $({\sigma}_\pi,~{\sigma}_K)=$ (30 mb,~15 mb), 
we obtain the differential lab cross section of $0^+_{\rm g.s.}$ in $^4_\Lambda$He as 
\begin{eqnarray}
  \left({d\sigma \over d\Omega_K}\right)^{J^\pi=0^+}_{\rm lab, 0^\circ}
  &=& 11.73 \,\mbox{$\mu$b/sr}.  \qquad \mbox{(DWIA)}
\label{eqn:e33}
\end{eqnarray}
We confirm that the differential lab cross sections in DWIA are relatively reduced 
by a distortion factor which is defined as  
\begin{eqnarray}
D_{\rm dis}& \equiv & N^{\rm DW}_{\rm eff}/N^{\rm PW}_{\rm eff}\simeq 0.3, 
\label{eqn:e34}
\end{eqnarray}
as shown in Table~\ref{tab:2}.
We find that the absolute values of the differential lab cross sections for SG (HO)
are larger (smaller) than those for Isle.
This is because the overlaps between wavefunctions between $N$ and $\Lambda$
are larger inside the nucleus in the order of SG, Isle, and HO, 
as seen in Fig.~\ref{fig:4}.

\begin{figure*}[t]
\begin{minipage}{0.49\linewidth}
\begin{center}
  \includegraphics[width=\linewidth]{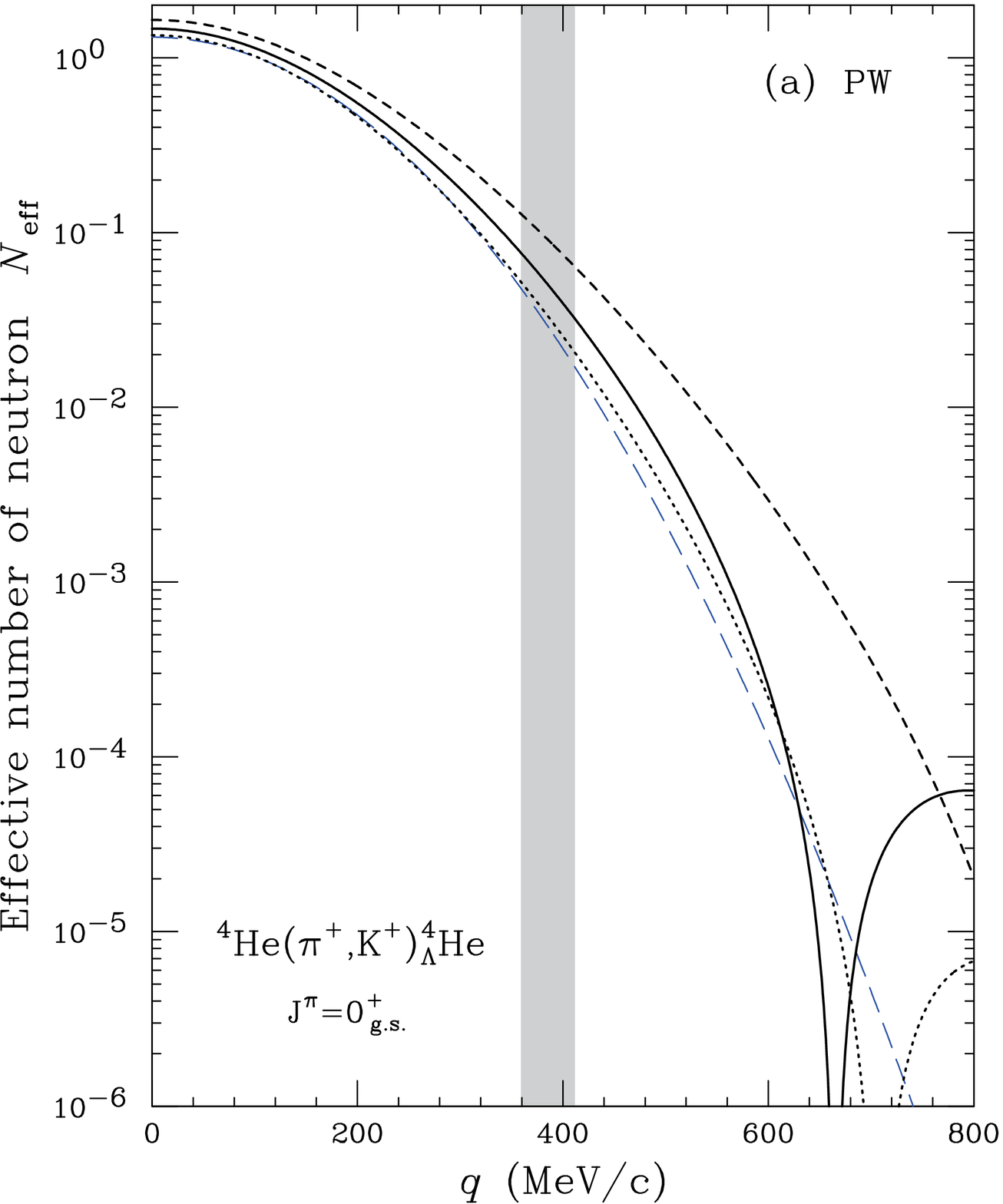}
\end{center}
\end{minipage}
\begin{minipage}{0.49\linewidth}
\begin{center}
  \includegraphics[width=\linewidth]{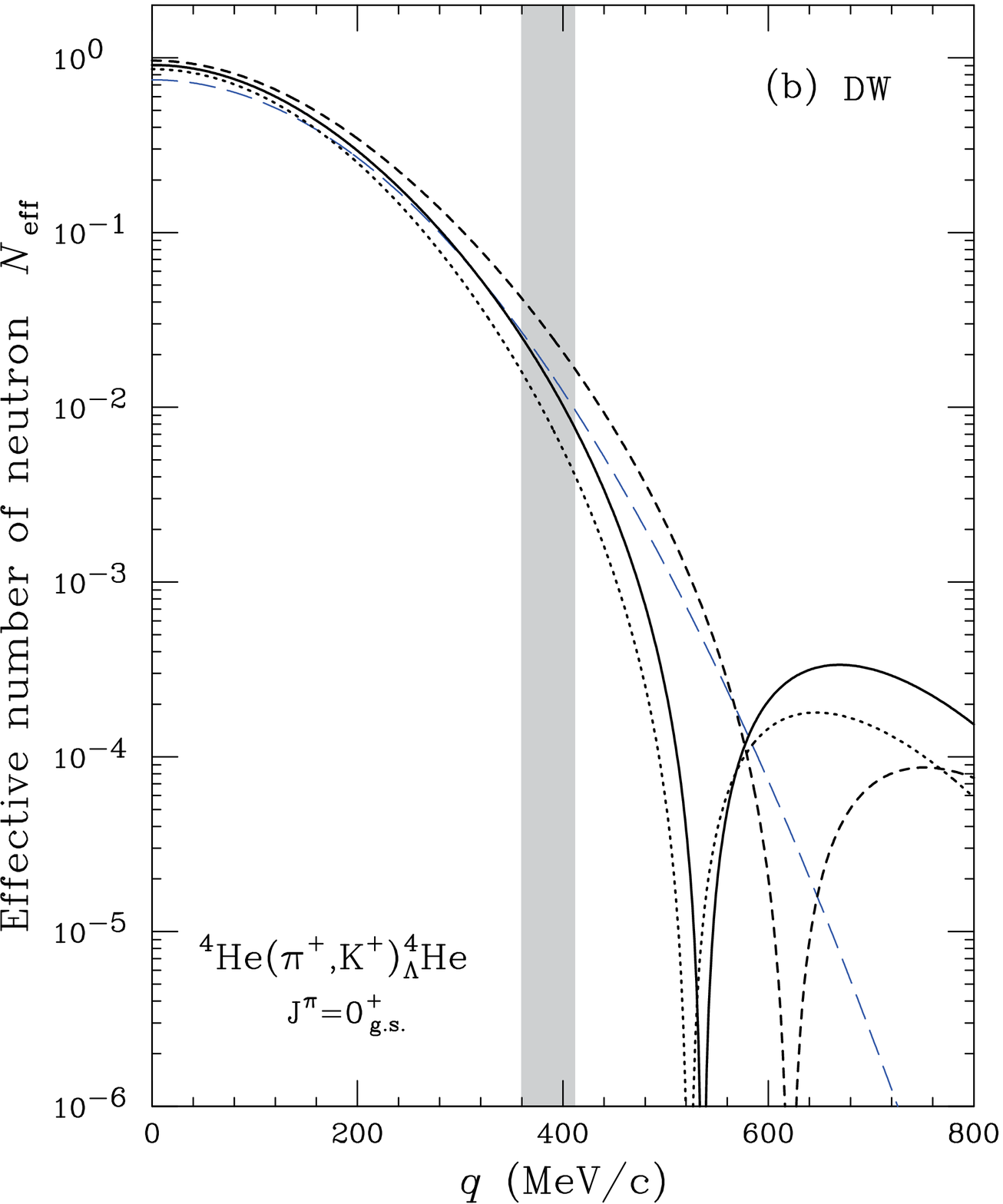}
\end{center}
\end{minipage}
\caption{\label{fig:6}
Calculated effective number of nucleons $N^{0^+}_{\rm eff}$ for $0^+_{\rm g.s.}$ 
in $^4_\Lambda$He via the $^4$He($\pi^+$,~$K^+$) reaction with (a) the plane-wave (PW) 
and (b) distorted-wave (DW) approximations, 
as a function of the momentum transfer $q$. 
Solid, dashed, and dotted curves denote the results obtained from 
the Isle, SG, and HO potentials, respectively. 
Long-dashed curves denote the results in the eikonal-oscillator approximation 
with $\bar{\sigma}=$ 30 mb. 
The halftones denote the region of $q=$ 360--410 MeV/$c$ corresponding to 
$\theta_{\rm lab}=$ $0^\circ$--$12^\circ$ at $p_\pi=$ 1.05 MeV/$c$.
}
\end{figure*}

\begin{table}[bt]
\caption{\label{tab:3}
Calculated PWIA and DWIA results of the differential lab cross sections of 
$0^+_{\rm g.s.}$ in $^4_\Lambda$He via the $^4$He($\pi^+$,~$K^+$) reaction at $p_{\pi}=$ 1.05 GeV/$c$.
The Isle, SG, and HO ($\hbar\omega_\Lambda=$ 7.0 MeV) potentials are used with 
$({\sigma}_\pi,~{\sigma}_K)=$ (30 mb,~15 mb).
The recoil effects are omitted.
}
\begin{ruledtabular}
\begin{tabular}{lccccccc}
\noalign{\smallskip}
& $\theta_{\rm lab}$
& {$N_{\rm eff}^{\rm PW}$}
& {$(d\sigma/d\Omega)_{\rm lab}^{\rm PW}$}
& {$D_{\rm dis}$}
& {$N_{\rm eff}^{\rm DW}$ }      
& {$(d\sigma/d\Omega)_{\rm lab}^{\rm DW}$} \\
& (degree)
& {($\times 10^{-2}$)}
& {($\mu$b/sr)} 
& 
& {($\times 10^{-2}$)} 
& {($\mu$b/sr)}   \\
\noalign{\smallskip}\hline\noalign{\smallskip}
\multicolumn{3}{l}{Isle}   \\
&0 & 0.886   & 4.113   & 0.093   & 0.082   & 0.382   \\
&4 & 0.756   & 3.684   & 0.077   & 0.058   & 0.283   \\
&8 & 0.466   & 2.181   & 0.034   & 0.016   & 0.074   \\
&12 & 0.199  & 0.829   & 0.000   & 0.000   & 0.000   \\
\noalign{\smallskip}
\multicolumn{3}{l}{SG}   \\
&0 & 2.413   & 11.23   & 0.162   & 0.391   &  1.813   \\
&4 & 2.154   & 10.59   & 0.151   & 0.325   &  1.583   \\
&8 & 1.538   & 7.572   & 0.119   & 0.183   &  0.857   \\
&12 & 0.886  & 3.683   & 0.071   & 0.063   &  0.263   \\
\noalign{\smallskip}
\multicolumn{5}{l}{HO}   \\
& 0  & 0.542  & 2.520 &  0.054  & 0.029 & 0.136 \\
& 4  & 0.463  &	2.256 &  0.039  & 0.018 & 0.092 \\
& 8  & 0.286  &	1.342 &  0.010  & 0.003 & 0.013 \\
&12  & 0.128  &	0.530 &  0.011  & 0.002 & 0.006 \\
\noalign{\smallskip}
\end{tabular}
\end{ruledtabular}
\end{table}

\subsubsection{$N^{0^+}_{\rm eff}$ v.s. $q$}

To see the features of the $\Lambda$ production of the nuclear ($\pi$, $K$) reactions, 
we study the effective number of nucleons $N_{\rm eff}^{J^\pi}$, 
as a function of the momentum transfer $q$ which is determined 
by the incident lab momentum $p_\pi$ and the $K$ forward-direction angles of $\theta_{\rm lab}$.

Figure~\ref{fig:6}(a) displays the calculated results of $N_{\rm eff}^{0^+}$, 
using the $\Lambda$ wavefunctions obtained from the Isle, SG and HO potentials.
The essential physical features of the endothermic nuclear ($\pi$, $K$) reactions 
can be well understood in terms of $N_{\rm eff}^{0^+}$ in PWIA, 
as suggested by Dover et al. \cite{Dover80}. 
We confirm that the magnitudes of $N_{\rm eff}^{0^+}$ obtained from the SG, Isle, and HO potentials 
in PWIA are large in this order. 
Note that the nuclear ($\pi$,~$K$) reactions at
$p_\pi=$ 1.05 GeV/$c$ ($\theta_{\rm lab}=$ $0^\circ$--$12^\circ$) provide
$q=$ 360--410 MeV/$c$ corresponding to the region of the halftones drawn in Fig.~\ref{fig:6}, 
where the recoil effects are very important owing to $q_{\rm eff}\simeq 0.75q$ in $^4_\Lambda$He. 
On the other hand, exothermic nuclear ($\bar{K}$, $\pi$) reactions with 
$p_\pi=$ 0.79 GeV/$c$ ($\theta_{\rm lab}=$ $0^\circ$--$12^\circ$) have  
$q=$ 68--126 MeV/$c$ which denote small momentum transfers, 
thus the recoil effects do not so affect the differential lab cross sections.
In the region of $q >$ 600 MeV/$c$, we find that the values of 
$N_{\rm eff}^{0^+}$ obtained from the Isle potential 
fall off, and their slopes are steeper than those with the SG potential;  
a dip appears in the region of $q=$ 600--800 MeV/$c$.
This behavior is well known to come from high momentum components in
the $\Lambda N$ and $NN$ wavefunctions due to short-range correlations, 
as discussed in Ref.~\cite{Shinmura86}.
We also find a dip in $N^{0^+}_{\rm eff}$ for HO which includes only the $NN$ correlations.

Figure~\ref{fig:6}(b) shows the calculated results of $N_{\rm eff}^{0^+}$ with DWIA. 
The distortions reduce the magnitudes of $N_{\rm eff}^{0^+}$ 
by about three times (see Table~\ref{tab:2}), changing slightly the shapes of $N_{\rm eff}^{0^+}$ 
in the region of $q <$ 400 MeV/$c$.
The slopes of $N_{\rm eff}^{0^+}$ become gradually steeper as increasing $q$ 
as in the case of $q >$ 400 MeV/$c$, and they grow a dip at the region of 
$q=$ 520--620 MeV/$c$ which can be achieved by $\theta_{\rm lab}=$ 24$^\circ$--34$^\circ$.
These behaviors may originate from high momentum components generated by meson distortions, 
because $D_0(r)$ significantly modifies $j_0(qr)$ inside the nucleus.
Therefore, the distortion effects as well as the recoil effects
are very important for large momentum transfer processes 
which can be realized in the ($\pi$, $K$) reactions.

\subsection{Comparison with eikonal-oscillator approximation}

We also consider the differential lab cross sections 
using the single-particle (s.p.) harmonic oscillator (HO) wavefunctions 
and the eikonal distortions by mesons, referring to it as 
the ``eikonal-oscillator'' approximation \cite{Dover80}.
This is often employed as nuclear model calculations 
for several reactions \cite{Dover83,Cieply01}. 
When we use the HO wavefunctions for both nucleon and $\Lambda$,  
we can express the eikonal distorted waves as 
\begin{eqnarray}
\chi_{K}^{(-)*}({\bm p}_K, {\bm r})\chi_{\pi}^{(+)}({\bm p}_\pi, {\bm r})
&=& \exp{(iqz)} \nonumber\\
&&\times \exp{\left(-\frac{\bar\sigma}{2}T({\bm b})\right)}
\label{eqn:e26}
\end{eqnarray}
at the $K$ forward direction angle of $\theta_{\rm lab}=$ 0$^\circ$; 
the nuclear thickness function for the $A=Z+N$ target nucleus is defined as
\begin{eqnarray}
T({\bm b})\equiv \int_{-\infty}^{\infty}\rho({\bm r})dz, \qquad 
\int T({\bm b})d{\bm b}=A, 
\label{eqn:e27}
\end{eqnarray}
with the averaged total cross section $\bar{\sigma}=(\sigma_\pi+\sigma_K)/2$
for the $\pi N$ and $K N$ elastic scatterings. 
Thus we have the effective number of nucleons $N^{J^\pi}_{\rm eff}$ 
for $0^+_{\rm g.s.}$ in Eq.~(\ref{eqn:e14}), 
which is rewritten as \cite{Dover80,Cieply01}
\begin{eqnarray}
N^{0^+}_{\rm eff}(\theta_{\rm lab})
&\simeq & 2\left({\bar{b}^6 \over \tilde{b}_\Lambda^3 \tilde{b}_N^3}\right)
  \exp{\left(-{1 \over 2}(\bar{b}\frac{M_C}{M_A}q)^2 \right)} \nonumber\\
&&\times |G_0(\bar{\sigma})|^2, 
\label{eqn:e28}
\end{eqnarray}
where the distorted-wave integral \cite{Dover80,Cieply01} is defined by 
\begin{eqnarray}
&&G_0(\bar{\sigma})=\int_0^\infty 2 t \exp(-t^2)
\exp\left( -\frac{\bar{\sigma}}{2}T(\bar{b}t)\right)dt, \nonumber\\
&&G_0(\bar{\sigma}=0)=1.
\label{eqn:e29}
\end{eqnarray}
Here the mean HO size parameter denotes
\begin{eqnarray}
1/\bar{b}^2=(1/\tilde{b}^2_\Lambda+1/\tilde{b}^2_N)/2.
\label{eqn:e30}
\end{eqnarray}
These formulas give us good insight for the nuclear ($\pi$, $K$) reactions.
The total effective number of nucleon $N_{\rm eff}^{\rm tot}$ 
at $\theta_{\rm lab}=$ $0^\circ$ is defined by the sum of 
all contributions of the $\Lambda$ final states.
In the closure approximation, it can be easily written as
\begin{eqnarray}
N_{\rm eff}^{\rm tot}(0^\circ)
&=&\sum_{J^\pi} N^{J^\pi}_{\rm eff}(0^\circ) \nonumber\\
&\simeq & \frac{N}{A}\int T({\bm b})\exp{\left(-\bar\sigma T({\bm b})\right)}d{\bm b}, 
\label{eqn:e30a}
\end{eqnarray}
so that the values of $N_{\rm eff}^{\rm tot}(0^\circ)$ amount to 
2.00, 1.22, and 0.99 for $\bar{\sigma}=$ 0 mb (PW), 20 mb, and 30 mb, respectively. 
Figure~\ref{fig:6} also displays the values of $N_{\rm eff}^{0^+}$ 
calculated in the eikonal-oscillator approximation of Eq.~(\ref{eqn:e28}). 
In the case of PWIA, the magnitude of $N^{0^+}_{\rm eff}$ is as large as 
that obtained from the HO potential, and it is smaller than that obtained from the Isle potential. 
In the case of DWIA ($\bar{\sigma}=$ 30 mb), 
the magnitude of $N^{0^+}_{\rm eff}$ is as large as that for Isle, 
and these slopes are similar to each other in the region of $q <$ 400 MeV/$c$.
Thus we obtain $(d\sigma/d\Omega)^{\rm DW}_{\rm lab}=$ 10.41, 9.83, 6.90, and 3.64 $\mu$b/sr
at $p_\pi=$ 1.05 GeV/$c$ in $\theta_{\rm lab}=$ 0$^\circ$, 4$^\circ$, 8$^\circ$, 
and 12$^\circ$, respectively; 
they are comparable to $(d\sigma/d\Omega)^{\rm DW}_{\rm lab}=$ 
11.73, 10.99, 7.47, and 3.65 $\mu$b/sr for Isle. 
Because high momentum components of neither the $N$ and $\Lambda$ wavefunctions 
nor the distorted waves for mesons are included 
in the eikonal-oscillator approximation, 
we confirm that the shape of $N^{0^+}_{\rm eff}$ has no dip as increasing $q$.

\subsection{\label{recoil}
Recoil effects}

\begin{figure}[t]
\begin{center}
  \includegraphics[width=\linewidth]{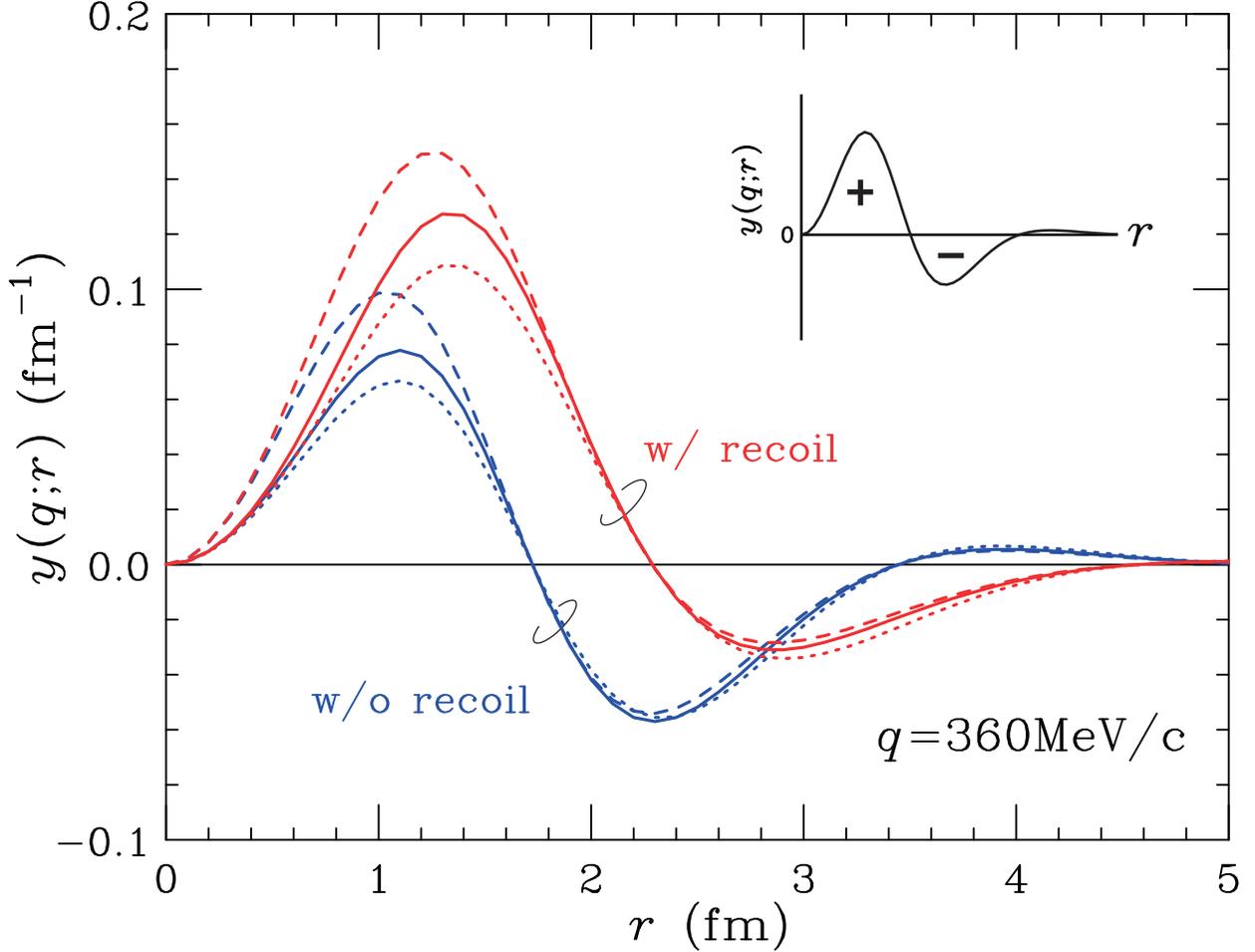}
\end{center}
\caption{\label{fig:7}
Overlap functions of ${\cal Y}(q; r)$ with/without the recoil effects 
in the $^4$He($\pi$,~$K$) reactions at $p_{\pi}=$ 1.05 GeV/$c$ (0$^\circ$)
which leads to  $q=$ 360 MeV, as a function of the radial distance. 
Solid and dashed, dotted curves denote the results obtained from 
the Isle, SG, and HO potentials, respectively. 
}
\end{figure}

The recoil effects should be needed in the ($\pi$,~$K$) reactions on 
a very light nuclear target such as $^4$He; 
the quantity of the recoil factor $M_C/M_A\simeq$ 3/4 = 0.75 characterizes 
the importance of the recoil effects in the nuclear systems. 
To see the sensitivity to the recoil effects quantitatively, 
we demonstrate the differential lab cross sections 
when we omit the recoil effects ($M_C/M_A \to 1$) in Eq.~(\ref{eqn:e7}) 
using the Isle, SG, and HO potentials.

In Table~\ref{tab:3}, we show the calculated DWIA (PWIA) results of 
the differential lab cross sections omitting the recoil effects
via the $^4$He($\pi$,~$K$) reaction at $p_{\pi}=$ 1.05 GeV/$c$.
We find $d\sigma/d\Omega_{\rm lab}=$ 0.382 $\mu$b/sr (4.113 $\mu$b/sr) 
at $\theta_{\rm lab}=$ 0$^\circ$. 
Surprisingly, this value is an order of magnitude smaller than 
$d\sigma/d\Omega_{\rm lab}=$ 11.73 $\mu$b/sr (35.14 $\mu$b/sr) 
which is already shown in Table~\ref{tab:2}.
The recoil effects have a great influence on $d\sigma/d\Omega_{\rm lab}$ 
depending on the radial behavior of the distorted waves for mesons.
Here we consider the overlap function defined as
\begin{eqnarray}
{\cal Y}(q; r)=r^2\rho_{00}^{(tr)}(r)\widetilde{j}_{0}(q;\frac{M_C}{M_A}r)
\label{eqn:e35}
\end{eqnarray}
which corresponds to the integrand in Eq.~(\ref{eqn:e17}).
Figure~\ref{fig:7} displays the behaviors of ${\cal Y}(q; r)$ 
for various types choosing the $\Lambda$ wavefunctions and the 
distortion parameters, as a function of the radial distance.
When we omit the recoil effects ($M_C/M_A \to 1$), 
we find that the node point at $r_1$ where ${\cal Y}(q; r_1)=0$ must 
be shifted toward the nuclear inside. 
As a result, the integral value of $\int_{0}^{\infty} {\cal Y}(q; r)dr=F(q)$
is significantly reduced by cancellation between the positive and negative values over 
integration regions in ${\cal Y}(q; r)$, as illustrated in Fig.~\ref{fig:7}. 

The recoil effects are often omitted in several model calculations 
for nuclear reactions with large momentum transfers, e.g., 
($\pi$,~$K$) \cite{Dover80}, ($\bar{K}$,~$K$) \cite{Dover83}, 
($\bar{K}$,~$p$) \cite{Cieply01}, and (stopped $\bar{K}$,~$\pi$) \cite{Matsuyama88} 
reactions on $^{12}$C in which the recoil effects are not so important 
because $M_C/M_A \simeq (A-1)/A$ = 11/12 = 0.917 for $A=$ 12.
But we must pay attention to the recoil effects when applying it 
to light nuclear systems such as $^{4}$He ($M_C/M_A \simeq $ 3/4 = 0.75). 
We believe that the calculated cross sections \cite{Dover83} 
or calculated production probabilities \cite{Matsuyama88} of $^4_\Lambda$He 
are perhaps underestimated by an order of magnitude for lack of the recoil effects \cite{Koike07}.

\subsection{Dependence on distortion parameters}

Due to strong absorptions of mesons in nuclei, e.g., 
$\pi^+$ at $p_\pi=$ 1.0--1.5 GeV/$c$ in the $N^*$ resonance region, 
the magnitude of the cross section may be also affected by meson distortions. 
To understand the distortion effects quantitatively, 
we demonstrate the differential lab cross sections of $0^+_{\rm g.s.}$ in $^4_\Lambda$He, 
considering various eikonal distortions with parameters of ($\sigma_\pi$, $\sigma_K$).
In Table~\ref{tab:4}, we show the calculated results of $d\sigma/d\Omega_{\rm lab}$ 
at $p_{\pi}=$ 1.05 GeV/$c$.
The magnitudes of the cross sections are reduced as increasing these parameters in DWIA. 
Thus the differential lab cross sections of $0^+_{\rm g.s.}$ in $^4_\Lambda$He 
at $\theta_{\rm lab}=$ 0$^\circ$ amount to 
$(d\sigma/d\Omega)_{\rm lab} \simeq$ 8--14 $\mu$b/sr which depend on $\pi$ and $K$ distorted waves 
for $\sigma_\pi=$ 20--30 mb and $\sigma_K=$ 10--30 mb.
It should be noticed that the parameter dependence 
gives an indication of the accuracy of our results 
within the eikonal distortion.
Fully realistic distorted waves obtained from
meson-nucleus optical potentials would be needed to make 
a more quantitative discussion. 

\begin{table}[tb]
\caption{\label{tab:4}
Distortion-parameter dependence of the differential lab cross sections 
of $0^+_{\rm g.s.}$ in $^4_\Lambda$He via the $^4$He($\pi^+$,~$K^+$) 
reaction at $p_{\pi}=$ 1.05 GeV/$c$. 
The Isle potential for $\Lambda$ is used.
}
\begin{ruledtabular}
\begin{tabular}{ccccc}
\noalign{\smallskip}
$({\sigma}_\pi,~{\sigma}_K)$
& $\theta_{\rm lab}$
& {$D_{\rm dis}$}
& {$N_{\rm eff}^{\rm DW}$ }
& {$(d\sigma/d\Omega)_{\rm lab}^{\rm DW}$} \\
(mb)
& (degree)
&
& {($\times 10^{-2}$)} 
& {($\mu$b/sr)}   \\
\noalign{\smallskip}\hline\noalign{\smallskip}
(20, 20)   & 0       & 0.387   & 2.932    & 13.61  \\     
           & 4       & 0.378   & 2.629    & 12.81   \\    
           & 8       & 0.351   & 1.889    & \ \ 8.85 \\   
           & 12      & 0.304   & 1.075    & \ \ 4.47 \\  
\noalign{\smallskip}                                                  
(30, 10)   & 0       & 0.372   & 2.818   & 13.08   \\    
           & 4       & 0.363   & 2.523   & 12.30   \\    
           & 8       & 0.335   & 1.805   & \ \ 8.45 \\   
           & 12      & 0.287   & 1.016   & \ \ 4.22 \\   
\noalign{\smallskip}                       
(30, 30)   & 0       & 0.243   & 1.841    & \ \ 8.54 \\   
           & 4       & 0.234   & 1.625    & \ \ 7.92 \\   
           & 8       & 0.206   & 1.110    & \ \ 5.20 \\   
           & 12      & 0.160   & 0.568    & \ \ 2.36 \\   
\end{tabular}
\end{ruledtabular}
\end{table}

\begin{figure}[bt]
\begin{center}
  \includegraphics[width=\linewidth]{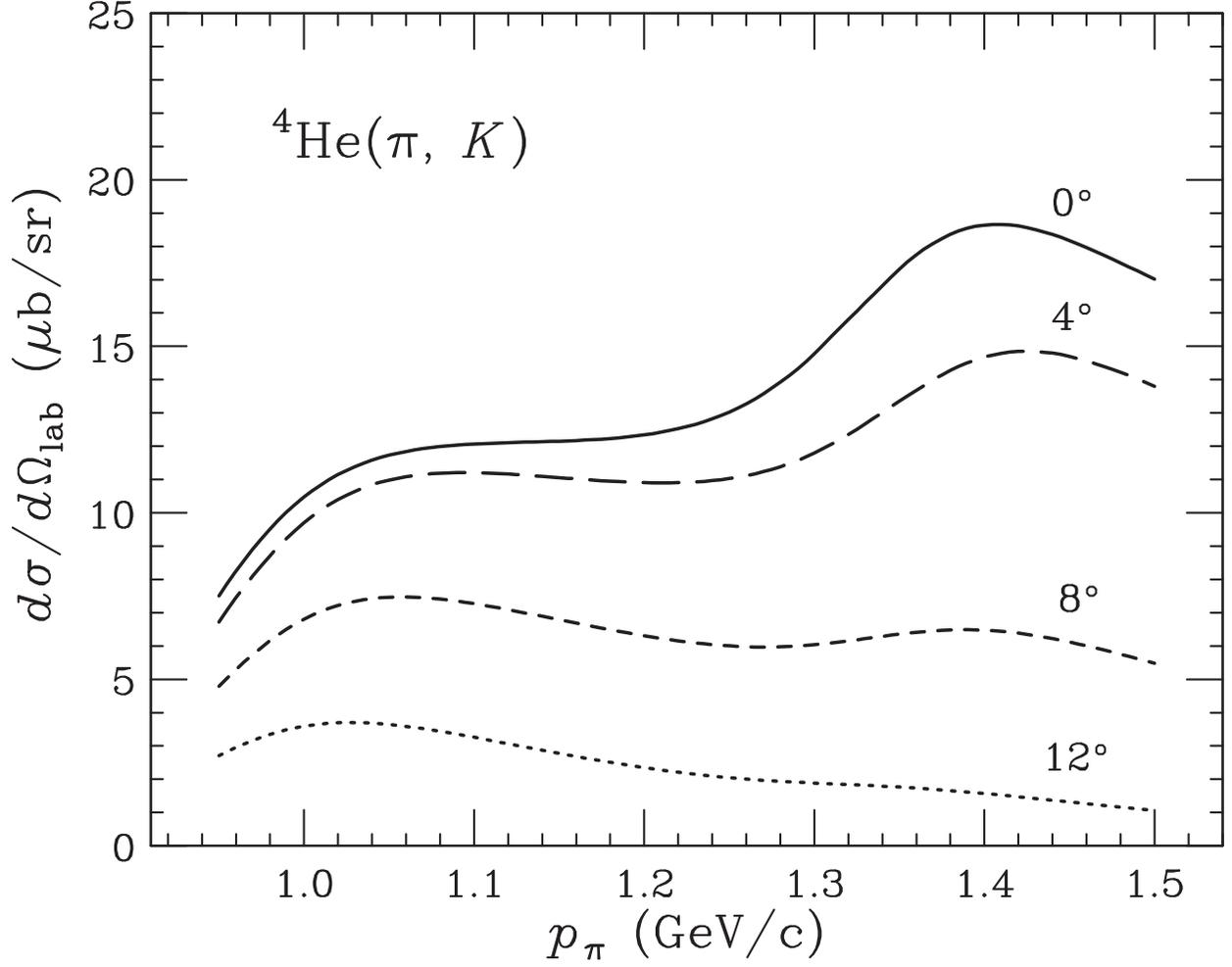}
\end{center}
\caption{\label{fig:8}
Incident-momentum dependence of the differential lab cross sections 
in the $^4$He($\pi$,~$K$) reactions at $\theta_{\rm lab}=$
0$^\circ$, 4$^\circ$, 8$^\circ$, and 12$^\circ$, as a function of $p_\pi$.
The $\Lambda$ wavefunctions obtained by the Isle potential 
and the distortion parameters of $({\sigma}_\pi,~{\sigma}_K)=$ (30 mb,~15 mb) are used. 
}
\end{figure}

\subsection{Dependence on the incident momentum}

In Fig.~\ref{fig:8}, we display the differential lab cross sections of $0^+_{\rm g.s.}$
in $^4_\Lambda$He via the $^4$He($\pi$,~$K$) reactions 
at $\theta_{\rm lab}=$ 0$^\circ$, 4$^\circ$, 8$^\circ$, and 12$^\circ$, 
as a function of the incident lab momentum $p_\pi$.
Here we used the $\Lambda$ wavefunctions obtained from the Isle potential 
and the eikonal distortions with $({\sigma}_\pi,~{\sigma}_K)=$ (30 mb,~15 mb). 
We find that the differential lab cross sections slightly increase, as increasing $p_\pi$. 
This trend seems to be opposite to that of 
$\alpha \langle d\sigma/d\Omega_{\rm lab} \rangle^{\rm opt}_{\pi N\to K\Lambda}$, 
as seen in Fig.~\ref{fig:2}.
This comes from the fact that the momentum transfers in this region
decrease as increasing $p_\pi$ (see Fig.~\ref{fig:1}), 
together with the nature of $N_{\rm eff}^{0^+}$ which must be taken 
into account the recoil effects.

\section{Summary and Conclusion}
\label{sect:6}

We have investigated theoretically the production cross sections of 
the $0^+$ ground state of a $^4_\Lambda$He hypernucleus in the $^4$He($\pi$,~$K$) reaction
with a distorted-wave impulse approximation using the optimal Fermi-averaged 
$\pi N\to K\Lambda$ $t$ matrix.  
We have demonstrated the sensitivity of the production cross section to
the $3N$-$\Lambda$ potentials and to the eikonal distorted waves for mesons. 
We have calculated the differential lab cross sections of $d\sigma/d\Omega_{\rm lab}$
at $p_{\pi}=$ 1.05 GeV/$c$ in the $K$ forward-direction angles of 
$\theta_{\rm lab}=$ 0$^\circ$--12$^\circ$.
The results can be summarized as follows:
\begin{itemize}
\item[(1)]  
The calculated differential lab cross section of $0^+_{\rm g.s.}$ in $^4_\Lambda$He 
amounts to $d\sigma/d\Omega_{\rm lab} \simeq$ 11 $\mu$b/sr at $p_{\pi}=$ 1.05 GeV/$c$, 
$\theta_{\rm lab}=$ 0$^\circ$--4$^\circ$, as in the case of the Isle potential.  

\item[(2)] 
The recoil effects enlarge the cross section of $^4_\Lambda$He via the 
$^4$He($\pi$, $K$) reactions by an order of magnitude, 
whereas the meson distortions reduce the cross section by 30\%. 

\item[(3)]
It is important to take into account the energy dependence of 
the $\pi N\to K\Lambda$ cross sections for a good description 
of the nuclear ($\pi$, $K$) reactions.

\item[(4)] 
The differential lab cross sections of $^4_\Lambda$H via $^4$He($\pi^-$, $K^0$) reactions  
are the same as those of $^4_\Lambda$He via $^4$He($\pi^+$, $K^+$) ones 
owing to charge independence in nuclear physics.

\end{itemize}
\noindent
In conclusion, we have shown that the differential lab cross sections of $0^+_{\rm g.s.}$ 
in $^4_\Lambda$He amount to $d\sigma/d\Omega_{\rm lab} \simeq$ 11 $\mu$b/sr at $p_{\pi}=$ 1.05 GeV/$c$ 
in the $K$ forward direction because of a major advantage of the use of the s-shell nuclear 
targets such as $^4$He.
It would be appropriate to study the $\Lambda$ production from the ($\pi$,~$K$) reactions 
on the $s$-shell $^{3,4}$He targets in order to study the lifetime measurements of
a $^{3,4}_\Lambda$H hypernucleus in production followed by mesonic decay processes. 
This investigation is now in progress \cite{Harada19}.

\begin{acknowledgments}
The authors would like to thank Professor A.~Sakaguchi, Professor H.~Tamura, 
and Dr. A.~Feliciello for many valuable discussions. 
This work was supported by Japan Society for
the Promotion of Science (JSPS), KAKENHI Grant Numbers JP16K05363.
\end{acknowledgments}

\appendix

\section{Explicit forms of isospin-spin functions for $0^+_{\rm g.s.}$ in $^4$He and $^4_\Lambda$He}
\label{app:1}

The isospin-spin function $X^A_{T_A,S_A}$ for $0^+_{\rm g.s.}$ in $^4$He
 ($T_A=0$, $m_{T_A}=0$; $S_A=0$, $m_{S_A}=0$) in Eq.~(\ref{eqn:e10}) 
is explicitly written as 
\begin{eqnarray}
&&X^A_{T_A,S_A}=
{\cal A}\bigl[\chi^{(3N)}_{I_3,S_3}\otimes \chi^{(N)}_{{1/2},{1/2}}\bigr]_{0,0} \nonumber\\
&& \quad
= \frac{1}{2\sqrt{6}}\big( 
-p_{\uparrow}  p_{\downarrow}n_{\uparrow}  n_{\downarrow}
+p_{\uparrow}  p_{\downarrow}n_{\downarrow}n_{\uparrow} 
+p_{\downarrow}p_{\uparrow}  n_{\uparrow}  n_{\downarrow} \nonumber\\
&& \quad
-p_{\downarrow}p_{\uparrow}  n_{\downarrow}n_{\uparrow}  
+p_{\uparrow}  n_{\uparrow}  p_{\downarrow}n_{\downarrow}
-p_{\uparrow}  n_{\downarrow}p_{\downarrow}n_{\uparrow} 
-p_{\downarrow}n_{\uparrow}  p_{\uparrow}  n_{\downarrow} \nonumber\\
&& \quad
+p_{\downarrow}n_{\downarrow}p_{\uparrow}  n_{\uparrow}  
-p_{\uparrow}  n_{\uparrow}  n_{\downarrow}p_{\downarrow}
+p_{\uparrow}  n_{\downarrow}n_{\downarrow}p_{\uparrow}
+p_{\downarrow}n_{\uparrow}  n_{\downarrow}p_{\uparrow}   \nonumber\\
&& \quad
-p_{\downarrow}n_{\downarrow}n_{\uparrow}  p_{\uparrow}
-n_{\uparrow}  p_{\uparrow}  p_{\downarrow}n_{\downarrow}
+n_{\uparrow}  p_{\downarrow}p_{\downarrow}n_{\uparrow}
+n_{\downarrow}p_{\uparrow}  p_{\downarrow}n_{\uparrow}   \nonumber\\
&& \quad
-n_{\downarrow}p_{\downarrow}p_{\uparrow}  n_{\uparrow}  
+n_{\uparrow}  p_{\uparrow}  n_{\downarrow}p_{\downarrow}
-n_{\uparrow}  p_{\downarrow}n_{\downarrow}p_{\uparrow}
-n_{\downarrow}p_{\uparrow}  n_{\uparrow}  p_{\downarrow}   \nonumber\\
&& \quad
+n_{\downarrow}p_{\downarrow}n_{\uparrow}  p_{\uparrow}  
-n_{\uparrow}  n_{\downarrow}p_{\uparrow}  p_{\downarrow}
+n_{\uparrow}  n_{\downarrow}p_{\downarrow}p_{\uparrow}
+n_{\downarrow}n_{\uparrow}  p_{\uparrow}  p_{\downarrow}   \nonumber\\
&& \quad
-n_{\downarrow}n_{\uparrow}  p_{\downarrow}p_{\uparrow}  \big),
\label{eqn:a1}
\end{eqnarray}
where $p_{\uparrow}$ and $n_{\uparrow}$ ($p_{\downarrow}$ and $n_{\downarrow}$) denote 
spin states with $m_z={+1/2}$ (${-1/2}$) for a proton and a neutron, respectively. 
The isospin-spin function $X^B_{T_B,S_B}$ for $0^+_{\rm g.s.}$ in $^4_\Lambda$He
($T_B=1/2$, $m_{T_B}=+1/2$; $S_B=0$, $m_{S_B}=0$) in Eq.~(\ref{eqn:e12}) 
is explicitly written as 
\begin{eqnarray}
&&X^B_{T_B,S_B}
=\bigl[\chi^{(3N)}_{I_3,S_3}\otimes \chi^{(\Lambda)}_{{0},{1/2}}\bigr]_{1/2,0}\nonumber\\
&& \quad
= \frac{1}{2\sqrt{3}}\big( 
-p_{\uparrow}  p_{\downarrow}n_{\uparrow}  \Lambda_{\downarrow}
+p_{\uparrow}  p_{\downarrow}n_{\downarrow}\Lambda_{\uparrow}
+p_{\downarrow}p_{\uparrow}  n_{\uparrow}  \Lambda_{\downarrow} \nonumber\\
&& \quad
-p_{\downarrow}p_{\uparrow}  n_{\downarrow}\Lambda_{\uparrow}
+p_{\uparrow}  n_{\uparrow}  p_{\downarrow}\Lambda_{\downarrow} 
+p_{\downarrow}n_{\downarrow}p_{\uparrow}  \Lambda_{\uparrow}
-n_{\uparrow}  p_{\uparrow}  p_{\downarrow}\Lambda_{\downarrow}  \nonumber\\
&& \quad
+n_{\uparrow}  p_{\downarrow}p_{\uparrow}  \Lambda_{\downarrow}
-n_{\uparrow}  p_{\downarrow}p_{\downarrow}\Lambda_{\uparrow} 
-n_{\downarrow}p_{\uparrow}  p_{\uparrow}  \Lambda_{\downarrow} 
+n_{\downarrow}p_{\uparrow}  p_{\downarrow}\Lambda_{\uparrow} \nonumber\\
&& \quad
-n_{\downarrow}p_{\uparrow}  p_{\downarrow}\Lambda_{\uparrow}\big),
\label{eqn:a2}
\end{eqnarray}
where $\Lambda_{\uparrow}$ ($\Lambda_{\downarrow}$) denotes a spin state with 
$m_z={+1/2}$ ($-1/2$) 
for a $\Lambda$ hyperon.


\clearpage

\end{document}